\documentclass[sn-mathphys,Numbered,iicol]{sn-jnl}

\usepackage{graphicx}%
\usepackage{multirow}%
\usepackage{amsmath,amssymb,amsfonts}%
\usepackage{amsthm}%
\usepackage{mathrsfs}%
\usepackage[title]{appendix}%
\usepackage{xcolor}%
\usepackage{textcomp}%
\usepackage{manyfoot}%
\usepackage{booktabs}%
\usepackage{algorithm}%
\usepackage{algorithmicx}%
\usepackage{algpseudocode}%
\usepackage{listings}%
\usepackage{fancyhdr}
\usepackage{csquotes}
\usepackage{bm}

\usepackage[utf8]{inputenc}  
\usepackage[T1]{fontenc}     
\usepackage{lmodern}         
\usepackage{microtype}       
\usepackage[scaled]{helvet}  
\usepackage[english]{babel}  
\usepackage{graphicx}        
\usepackage{xspace}

\theoremstyle{thmstyleone}%
\theoremstyle{thmstyletwo}%

\theoremstyle{thmstylethree}%

\begin{document}

%

\newcommand{\pp}           {\ensuremath{\mathrm{pp}}\xspace}
\newcommand{\ppbar}        {\mbox{\ensuremath{\mathrm {p\overline{p}}}}\xspace}
\newcommand{\XeXe}         {\ensuremath{\mathrm{\mbox{XeXe}}}\xspace}
\newcommand{\PbPb}         {\ensuremath{\mathrm{\mbox{PbPb}}}\xspace}
\newcommand{\pA}           {\ensuremath{\mathrm{\mbox{pA}}}\xspace}
\newcommand{\pPb}          {\ensuremath{\mathrm{\mbox{pPb}}}\xspace}
\newcommand{\AuAu}         {\ensuremath{\mathrm{\mbox{AuAu}}}\xspace}
\newcommand{\OO}         {\ensuremath{\mathrm{\mbox{OO}}}\xspace}
\renewcommand{\AA}           {\ensuremath{\mathrm{\mbox{AA}}}\xspace}
\newcommand{\dAu}          {\ensuremath{\mathrm{\mbox{dAu}}}\xspace}
\newcommand{\ee}           {\ensuremath{\mathrm{e^{+}e^{-}}}\xspace}
\newcommand{\ep}           {\ensuremath{\mathrm{e^{\pm}p\xspace}}}
\newcommand{\qqbar}        {\ensuremath{\mathrm{q}\overline{\mathrm{q}}}\xspace}

\newcommand{\s}            {\ensuremath{\sqrt{s}}\xspace}
\newcommand{\snn}          {\ensuremath{\sqrt{s_{\mathrm{NN}}}}\xspace}
\newcommand{\pt}           {\ensuremath{p_{\rm T}}\xspace}
\newcommand{\meanpt}       {$\langle p_{\mathrm{T}}\rangle$\xspace}
\newcommand{\ycms}         {\ensuremath{y_{\rm CMS}}\xspace}
\newcommand{\ylab}         {\ensuremath{y_{\rm lab}}\xspace}
\newcommand{\etarange}[1]  {\mbox{$\left | \eta \right |~<~#1$}}
\newcommand{\yrange}[1]    {\mbox{$\left | y \right |~<~#1$}}
\newcommand{\dndy}         {\ensuremath{\mathrm{d}N_\mathrm{ch}/\mathrm{d}y}\xspace}
\newcommand{\dndeta}       {\ensuremath{\mathrm{d}N_\mathrm{ch}/\mathrm{d}\eta}\xspace}
\newcommand{\avdndeta}     {\ensuremath{\langle\dndeta\rangle}\xspace}
\newcommand{\dNdy}         {\ensuremath{\mathrm{d}N_\mathrm{ch}/\mathrm{d}y}\xspace}
\newcommand{\Npart}        {\ensuremath{N_\mathrm{part}}\xspace}
\newcommand{\Ncoll}        {\ensuremath{N_\mathrm{coll}}\xspace}
\newcommand{\dEdx}         {\ensuremath{\textrm{d}E/\textrm{d}x}\xspace}
\newcommand{\RpPb}         {\ensuremath{R_{\rm pPb}}\xspace}
\newcommand{\Tfo}         {\ensuremath{T_\mathrm{fo}}\xspace}
\newcommand{\Th}         {\ensuremath{T_\mathrm{h}}\xspace}
\newcommand{\de}          {\ensuremath{\mathrm{d}}\xspace}
\newcommand{\Raa}          {\ensuremath{R_\mathrm{AA}}\xspace}
\newcommand{\FFc}{\ensuremath{f(\mathrm{c\rightarrow H_{c}})}\xspace}
\newcommand{\zjet}{\ensuremath{z_{\mathrm{||}}^{\mathrm{ch}}}\xspace}
\newcommand{\nineH}        {$\sqrt{s}~=~0.9$~Te\kern-.1emV\xspace}
\newcommand{\seven}        {$\sqrt{s}~=~7$~Te\kern-.1emV\xspace}
\newcommand{\twoH}         {$\sqrt{s}~=~0.2$~Te\kern-.1emV\xspace}
\newcommand{\twosevensix}  {$\sqrt{s}~=~2.76$~Te\kern-.1emV\xspace}
\newcommand{\five}         {$\sqrt{s}~=~5.02$~Te\kern-.1emV\xspace}
\newcommand{\twosevensixnn}{$\sqrt{s_{\mathrm{NN}}}~=~2.76$~Te\kern-.1emV\xspace}
\newcommand{\fivenn}       {$\sqrt{s_{\mathrm{NN}}}~=~5.02$~Te\kern-.1emV\xspace}
\newcommand{\LT}           {L{\'e}vy-Tsallis\xspace}
\newcommand{\GeVc}         {\ensuremath{\mathrm{GeV}/c}\xspace}
\newcommand{\MeVc}         {\ensuremath{\mathrm{MeV}/c}\xspace}
\newcommand{\TeV}          {\ensuremath{\mathrm{TeV}}\xspace}
\newcommand{\GeV}          {\ensuremath{\mathrm{GeV}}\xspace}
\newcommand{\MeV}          {\ensuremath{\mathrm{MeV}}\xspace}
\newcommand{\GeVmass}      {\ensuremath{\mathrm{GeV}/c^2}\xspace}
\newcommand{\MeVmass}      {\ensuremath{\mathrm{MeV}/c^2}\xspace}
\newcommand{\lumi}         {\ensuremath{\mathcal{L}}\xspace}
\newcommand{\mub}         {\ensuremath{\mu\mathrm{b}}\xspace}
\newcommand{\Taa}          {\ensuremath{\langle T_\mathrm{AA}\rangle}\xspace}
\newcommand{\Diffs}        {\ensuremath{D_{s}}\xspace}
\newcommand{\vtwo}          {\ensuremath{v_\mathrm{2}}\xspace}
\newcommand{\ITS}          {\rm{ITS}\xspace}
\newcommand{\TOF}          {\rm{TOF}\xspace}
\newcommand{\ZDC}          {\rm{ZDC}\xspace}
\newcommand{\ZDCs}         {\rm{ZDCs}\xspace}
\newcommand{\ZNA}          {\rm{ZNA}\xspace}
\newcommand{\ZNC}          {\rm{ZNC}\xspace}
\newcommand{\SPD}          {\rm{SPD}\xspace}
\newcommand{\SDD}          {\rm{SDD}\xspace}
\newcommand{\SSD}          {\rm{SSD}\xspace}
\newcommand{\TPC}          {\rm{TPC}\xspace}
\newcommand{\TRD}          {\rm{TRD}\xspace}
\newcommand{\VZERO}        {\rm{V0}\xspace}
\newcommand{\VZEROA}       {\rm{V0A}\xspace}
\newcommand{\VZEROC}       {\rm{V0C}\xspace}
\newcommand{\Vdecay} 	   {\ensuremath{V^{0}}\xspace}

\newcommand{\pip}          {\ensuremath{\pi^{+}}\xspace}
\newcommand{\pim}          {\ensuremath{\pi^{-}}\xspace}
\newcommand{\kap}          {\ensuremath{\rm{K}^{+}}\xspace}
\newcommand{\kam}          {\ensuremath{\rm{K}^{-}}\xspace}
\newcommand{\pbar}         {\ensuremath{\rm\overline{p}}\xspace}
\newcommand{\kzero}        {\ensuremath{{\rm K}^{0}_{\rm{S}}}\xspace}
\newcommand{\lmb}          {\ensuremath{\Lambda}\xspace}
\newcommand{\almb}         {\ensuremath{\overline{\Lambda}}\xspace}
\newcommand{\Om}           {\ensuremath{\Omega^-}\xspace}
\newcommand{\Mo}           {\ensuremath{\overline{\Omega}^+}\xspace}
\newcommand{\X}            {\ensuremath{\Xi^-}\xspace}
\newcommand{\Ix}           {\ensuremath{\overline{\Xi}^+}\xspace}
\newcommand{\Xis}          {\ensuremath{\Xi^{\pm}}\xspace}
\newcommand{\Oms}          {\ensuremath{\Omega^{\pm}}\xspace}
\newcommand{\degree}       {\ensuremath{^{\rm o}}\xspace}
\newcommand{\Hc}           {\ensuremath{\mathrm{H_c}}\xspace}
\newcommand{\Dzero}        {\ensuremath{\mathrm{D^0}}\xspace}
\newcommand{\Dplus}        {\ensuremath{\mathrm{D^+}}\xspace}
\newcommand{\Dminus}        {\ensuremath{\mathrm{D^-}}\xspace}

\newcommand{\Dstar}        {\ensuremath{\mathrm{D^{*+}}}\xspace}
\newcommand{\Dstarzero}        {\ensuremath{\mathrm{D^{*0}}}\xspace}
\newcommand{\Ds}           {\ensuremath{\mathrm{D_s^+}}\xspace}
 \newcommand{\Bs}           {\ensuremath{\mathrm{B_s^0}}\xspace}
 \newcommand{\Bplus}           {\ensuremath{\mathrm{B^+}}\xspace}
  \newcommand{\Bzero}           {\ensuremath{\mathrm{B^0}}\xspace}
  \newcommand{\Bc}           {\ensuremath{\mathrm{B^+_c}}\xspace}

\newcommand{\Dsstar}{\ensuremath{\mathrm{D_s^{*+}}}\xspace}

\newcommand{\Dsminus}    
{\ensuremath{\mathrm{D_s^-}}\xspace}
\newcommand{\Lc}           {\ensuremath{\Lambda_\mathrm{c}^+}\xspace}
\newcommand{\Lcgeneric}           {\ensuremath{\Lambda_\mathrm{c}}\xspace}
\newcommand{\Lcminus}           {\ensuremath{\Lambda_\mathrm{c}^-}\xspace}
\newcommand{\LcExcitedOne}{\ensuremath{\Lambda_\mathrm{c}(2595)^+}\xspace}
\newcommand{\LcExcitedTwo}{\ensuremath{\Lambda_\mathrm{c}(2625)^+}\xspace}
\newcommand{\SigmacZero}           {\ensuremath{\Sigma_\mathrm{c}(2455)^{0}}\xspace}
\newcommand{\SigmacZeroPlus}           {\ensuremath{\Sigma_\mathrm{c}(2455)^{0,++}}\xspace}
\newcommand{\SigmacZeroExcited}           {\ensuremath{\Sigma_\mathrm{c}(2520)^{0}}\xspace}

\newcommand{\Sigmac}           {\ensuremath{\Sigma_\mathrm{c}^{0,+,++}}\xspace}
\newcommand{\Sigmacgeneric}           {\ensuremath{\Sigma_\mathrm{c}}\xspace}

\newcommand{\XicZero}      {\ensuremath{\Xi_\mathrm{c}^0}\xspace}
\newcommand{\XicPlus}      {\ensuremath{\Xi_\mathrm{c}^+}\xspace}
\newcommand{\XicPlusZero}  {\ensuremath{\Xi_\mathrm{c}^{0,+}}\xspace}
\newcommand{\Omegac}       {\ensuremath{\Omega_\mathrm{c}^0}\xspace}
\newcommand{\Jpsi}         {\ensuremath{\mathrm{J}/\psi}\xspace}
\newcommand{\PsiTwos}         {\ensuremath{\psi(\mathrm{2S})}\xspace}
\newcommand{\UpsilonOneS}         {\ensuremath{\mathrm{\Upsilon (1S)}}\xspace}
\newcommand{\ccbar}        {\ensuremath{\mathrm{c\overline{c}}}\xspace}
\newcommand{\bbbar}        {\ensuremath{\mathrm{b\overline{b}}}\xspace}

\newcommand{\Xib}          {\ensuremath{\Xi_\mathrm{b}^{0,-}}\xspace}
\newcommand{\DzerotoKpi}   {\ensuremath{\mathrm{D^0\to K^-\pi^+}}}
\newcommand{\Lb}     {\ensuremath{\Lambda_\mathrm{b}^0}\xspace}
\newcommand{\Lbbar}           {\ensuremath{\bar{\Lambda}_\mathrm{b}^0}\xspace}

\newcommand{\lowptbin}{\ensuremath{0<\pt<1}~\GeVc}

\newcommand{\LcD} {\ensuremath{\Lc/\Dzero}\xspace}
\newcommand{\QQbar}           {\ensuremath{\mathrm{Q\bar{Q}}\xspace}}

\newcommand{\tamu}         {\textsc{tamu}\xspace}
\newcommand{\pythiasix}    {\textsc{pythia6}\xspace}
\newcommand{\herwig}    {\textsc{herwig}\xspace}
\newcommand{\pythiaeight}  {\textsc{pythia8}\xspace}
\newcommand{\pythiaeightprecise}{\textsc{pythia8.243}\xspace}
\newcommand{\pythiasixprecise}{\textsc{pythia6.4.25}\xspace}
\newcommand{\pythia}       {\textsc{pythia}\xspace}
\newcommand{\hijing}       {\textsc{hijing}\xspace}
\newcommand{\hijingprecise}{\textsc{hijing v1.383}\xspace}
\newcommand{\fonll}        {\textsc{fonll}\xspace}
\newcommand{\evtgen}       {\textsc{EvtGen}\xspace}

\title[Article Title]{Aspects of Higgs Physics at a $\sqrt{s}=3$ TeV Muon Collider with detailed detector simulation}

\author[a]{Paolo Andreetto}
\author[b]{Nazar Bartosik}
\author[a,c]{Laura Buonincontri}
\author[c,d]{Daniele Calzolari}
\author[e,f]{Vieri Candelise}
\author[e]{Massimo Casarsa}
\author[g,h]{Luca Castelli}
\author[i]{Mauro Chiesa}
\author[j,k]{Anna Colaleo}
\author[l]{Giacomo Da Molin}
\author[m]{Matthew Forslund}
\author[a,c]{Luca Giambastiani}
\author[a]{Alessio Gianelle}
\author[e,f]{Carlo Giraldin}
\author[n]{Karol Krizka}
\author[o]{Sergo Jindariani}
\author[d]{Anton Lechner}
\author[a,c]{Donatella Lucchesi}
\author[c]{Leo Mareso}
\author[p]{Paola Mastrapasqua}
\author[m]{Patrick Meade}
\author[q]{Alessandro Montella}
\author[r]{Simone Pagan Griso}
\author[a,c]{Leonardo Palombini}
\author[b]{Nadia Pastrone}
\author[s]{Lorenzo Sestini}
\author[j,k]{Rosamaria Venditti}
\author[j,k]{Angela Zaza}
\author[a,c]{Davide Zuliani}

\affil[a]{INFN Sezione di Padova, Padova, Italy}
\affil[b]{INFN Sezione di Torino, Torino, Italy}
\affil[c]{Università di Padova, Padova, Italy}
\affil[d]{European Organization for Nuclear Research, Geneva, Switzerland}
\affil[e]{INFN Sezione di Trieste, Trieste, Italy}
\affil[f]{Università di Trieste, Trieste, Italy}
\affil[g]{INFN Sezione di Roma, Roma, Italy}
\affil[h]{Università La Sapienza, Roma, Italy}
\affil[i]{INFN Sezione di Pavia, Pavia, Italy}
\affil[j]{INFN Sezione di Bari, Bari, Italy}
\affil[k]{Università di Bari, Bari, Italy}
\affil[l]{Laboratório de Instrumentação e Física Experimental de Partículas, Lisboa, Portugal}
\affil[m]{Stony Brook University, Stony Brook, United States}
\affil[n]{University of Birmingham, Birmingham, United Kingdom}
\affil[o]{Fermi National Accelerator Laboratory, Batavia, United States}
\affil[p]{Université Catholique de Louvain, Louvain-la-Neuve, Belgium}
\affil[q]{Stockholms Universitet, Stockholm, Sweden}
\affil[r]{Lawrence Berkeley National Laboratory, Berkeley, United States}
\affil[s]{INFN Sezione di Firenze, Firenze, Italy}
\affil[]{\newline \newline \newline \small Contacts: Massimo Casarsa (\href{massimo.casarsa@ts.infn.it}{massimo.casarsa@ts.infn.it}) and Lorenzo Sestini (\href{lorenzo.sestini@cern.ch}{lorenzo.sestini@cern.ch})}


\abstract{The Muon Collider is one of the most promising future collider facilities with the potential to reach multi-TeV center-of-mass energy and high luminosity. Due to the significant Higgs boson production cross section in muon-antimuon collisions at such high energies, the collider offers an excellent opportunity for in-depth exploration of Higgs boson properties. It holds the capability to significantly advance our understanding of the Higgs sector to a very high level of precision. However, the presence of beam-induced background resulting from the decay of the beam muons poses unique challenges for detector development and event reconstruction. In this paper, the prospects for measuring various Higgs boson properties at a center-of-mass energy of 3 TeV are presented, using a detailed detector simulation in a realistic environment. The study demonstrates the feasibility of achieving high precision results  with the current state-of-the-art detector design. In addition, the paper discusses the detector requirements  necessary to achieve this level of accuracy. }

\maketitle

\section{Introduction}
\label{sec:intro}

\begin{figure*}[!h]
\centering
\includegraphics[width=0.48\textwidth]{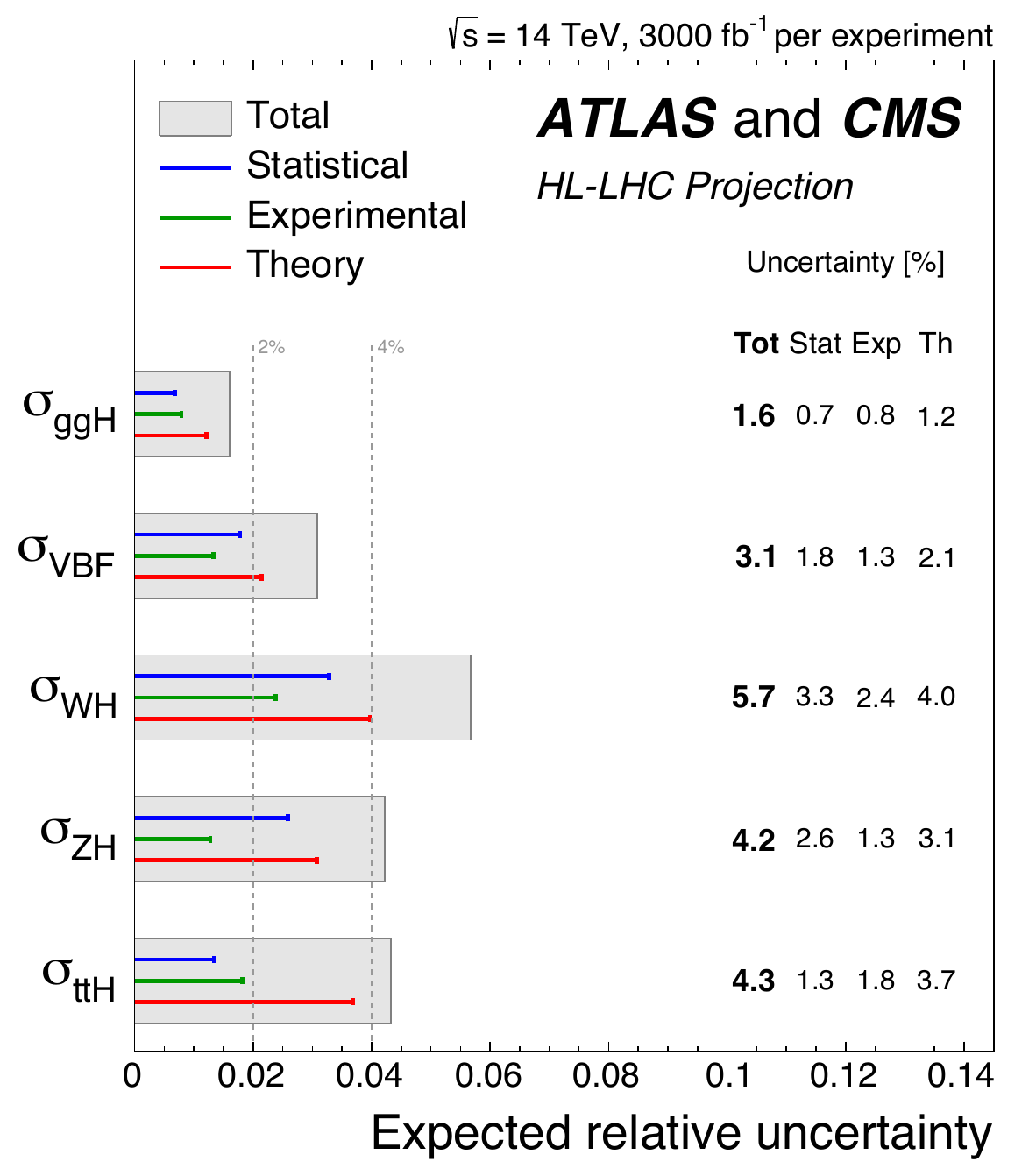}\hspace{0.3cm}
\includegraphics[width=0.48\textwidth]{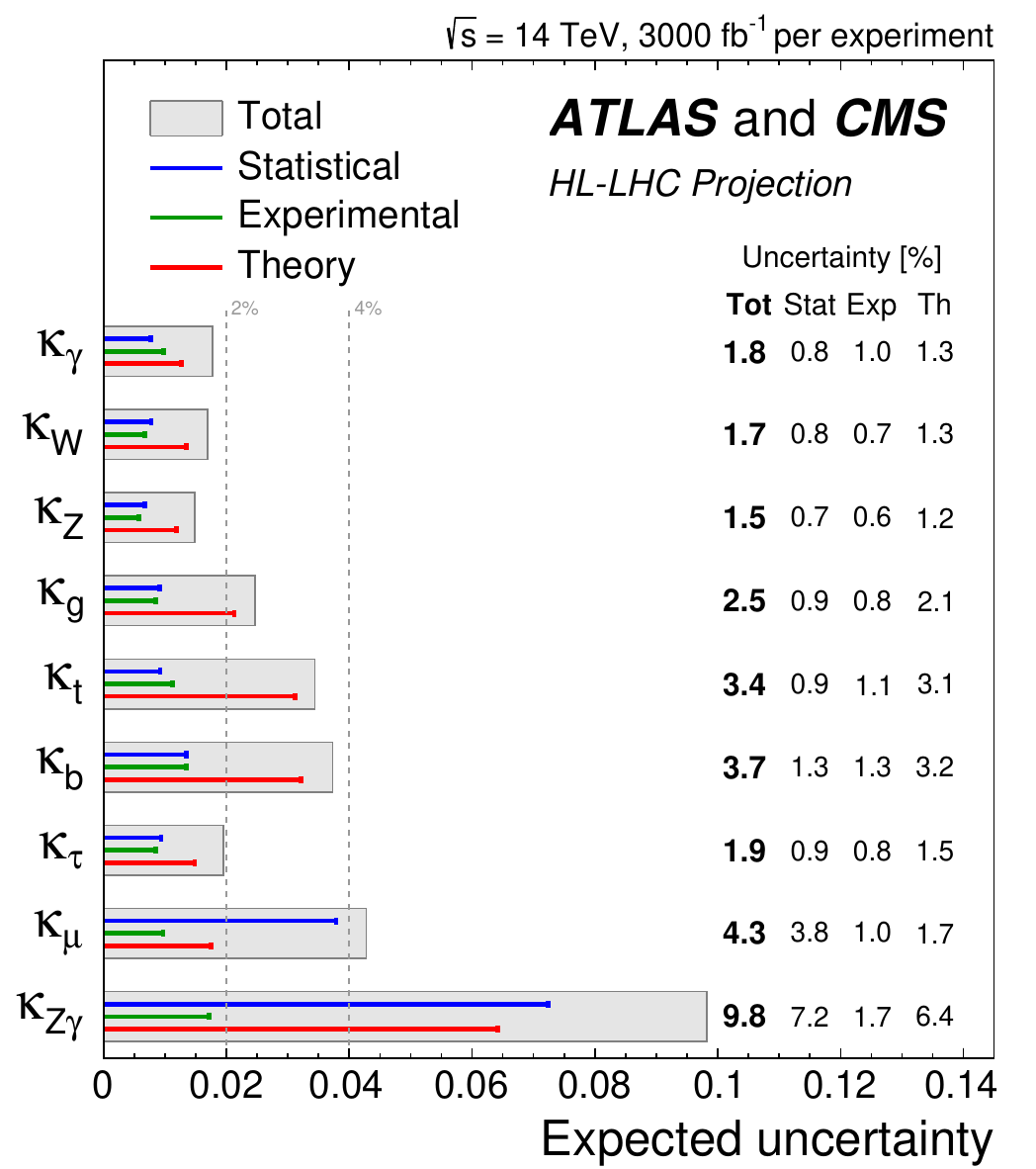}
\caption{Projections of the expected precision of the ATLAS and CMS combination on the Higgs production cross sections (left) and $\kappa_i$ coupling modifiers (right) including HL-LHC data. These figures have been taken from another publication \cite{hllhc_snowmass}, dated back to 2022.}
\label{fig:higgsHLLHC}
\end{figure*}

The Higgs boson ($H$) is considered a portal to new physics, because it is connected to some of the fundamental questions about the Universe~\cite{higgs_report}, including the mechanism of Electroweak Symmetry Breaking (EWSB), the origin of the masses, the matter-antimatter asymmetry, and the nature of dark matter. The EWSB \cite{higgs1,higgs2,higgs3,higgs4} is formulated via the scalar potential, which is written below in a form that includes possible deviations from the Standard Model (SM):
\begin{equation}
    V(h)=\frac{1}{2} m_H^2 h^2+\lambda_3 vh^3+\frac{1}{4} \lambda_4 h^4 + \mathcal{O}(h^4) ,
    \label{eq:higgspotential}
\end{equation}
where $h$ represents the field of the physical Higgs boson, which is what remains after the EWSB.
The Higgs boson mass is $m_H = \sqrt{2}\lambda v = 125.20 \pm 0.11$ GeV\footnote{In this paper natural units where $\hbar=c=1$ are used.}~\cite{ref:pdg},  $v=1/\sqrt{\sqrt{2} G_F } \simeq 246 $ GeV is the vacuum expectation value, and $G_F$ is the Fermi constant. In the SM, $\lambda_3=\lambda_4=\lambda=m_H^2/(2\,v^2)$ represents the strength of the coupling of the Higgs boson to itself. 
The couplings of the Higgs boson to the elementary fermions and bosons generate the particle masses. 
The most recent review of the expected experimental precision on such couplings, including projections for the High-Luminosity LHC (HL-LHC), is presented in the Snowmass Energy Frontier report~\cite{snowmassEF}. Figure~\ref{fig:higgsHLLHC} summarises the expected precision on the Higgs production cross sections and coupling modifiers $\kappa_{i}$ \cite{kframework} with 3000 fb$^{-1}$ of data that will be collected by ATLAS and CMS at the end of HL-LHC.

In the case of the Higgs self-interaction, the 95\% C.L. limits on the coupling modifier $\kappa_{\lambda_3}=\lambda_3/\lambda_3^{\mathrm{SM}}$ (with $\lambda_3^{\mathrm{SM}}$ being the expected SM value) are $- 0.6 < \kappa_{\lambda_3} < 6.6$ for ATLAS \cite{atlas_lambda} and $- 1.24 < \kappa_{\lambda_3} < 6.49$ for CMS \cite{cms_lambda} with the current LHC data. With HL-LHC data, a 50\% precision on the SM value of the self-coupling is expected~\cite{higgs_report}, which is not sufficient to determine the potential shape.

In order to reveal new physics in the Higgs sector, it is essential to determine the Higgs couplings to elementary fermions and bosons with a percent-level precision and achieve a few percent precision in the measurement of the Higgs self-coupling. However, it should be noted that the required precision may vary depending on the specific new-physics scenario under consideration. This underscores the need for a dedicated Higgs factory.

Linear and circular $e^+e^-$ machines have been the subject of debate for a long time as well as $\mu^+\mu^-$ colliders. Recently, $\mu^+\mu^-$ collisions have been regaining attention on the international scene, thanks to their impressive physics potential and new advancements on the technical feasibility~\cite{epjc}.

A Muon Collider (MuC) is a proposed future accelerator in which $\mu^+$ and $\mu^-$ beams collide at very high center-of-mass energies in a circular machine, thanks to the negligible energy loss of muons due to synchrotron radiation. The current proposal of the International Muon Collider Collaboration (IMCC)~\cite{imcc} aims at 10 TeV center-of-mass energy collisions, with a possible initial phase at 3 TeV~\cite{epjc}.
A comprehensive review of the physics potential of MuC experiments at both $\sqrt{s}=3$ and $\sqrt{s}=10$ TeV can be found in Refs.~\cite{musmasher,epjc,3tevphysics}. 

The determination of the Higgs couplings and self-coupling with an unprecedented accuracy will be feasible with 10 TeV center-of-mass muon-antimuon collisions. However, given the Higgs boson production cross section and the achievable machine luminosity,
the possibility of a run at $\sqrt{s}=3$ TeV would be a valuable option for Higgs boson research. 

This paper presents results on Higgs boson studies obtained at a $\sqrt{s}=3$ TeV MuC using a detailed detector simulation that includes the beam-induced background (BIB).
The BIB poses a significant challenge to extracting physics in a MuC experiment. However, as will be shown in the next sections, the precision achieved with the detailed detector simulation with the BIB is comparable to the detector target performance estimated without taking into account the BIB effects.


The experimental facility is introduced in Sec.~\ref{sec:facility}, focusing on the interaction region and the detector shielding. The paper, then, describes 
the generation of physics and background samples, the detector model and the detector simulation, and the event reconstruction in the presence of the BIB (Sec.~\ref{sec:reconstruction}). The results obtained for the Higgs boson production cross sections times the branching ratio for several decay channels are discussed in detail in Secs.~\ref{sec:h2bb}-\ref{sec:h2gammagamma}, and are compared to the target simulation studies in Sec.~\ref{sec:parametric}. 
In Sec.~\ref{sec:width}, the potential of a muon collider for measuring the Higgs boson width is investigated.
The double Higgs production cross section and the self-coupling precision determination are presented in Secs.~\ref{sec:HH} and \ref{sec:trilinear}, respectively.
A discussion on systematic uncertainties is presented in Sec.~\ref{sec:sys}.
The lessons learned on the detector requirements necessary for high-precision Higgs physics measurements are reported in Sec.~\ref{sec:requirements}. Finally, Sec.~\ref{sec:conclusions} summarizes the prospects for Higgs physics in a MuC experiment.

\section{Experimental environment}
\label{sec:facility}

\begin{figure*}[th!]
\centering
\includegraphics[width=0.495\textwidth]{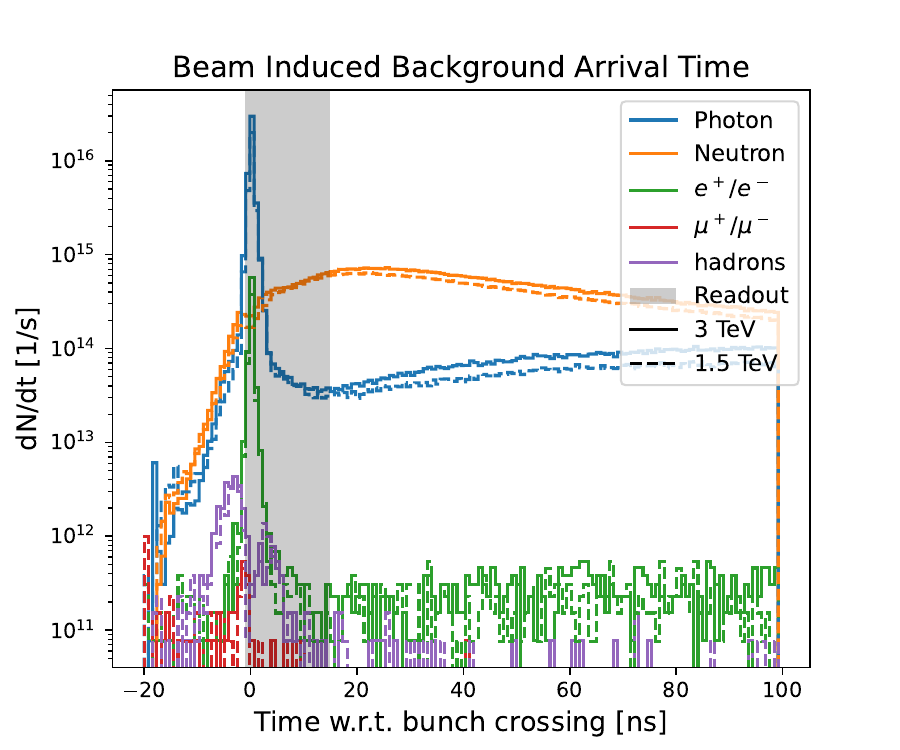}
\includegraphics[width=0.495\textwidth]{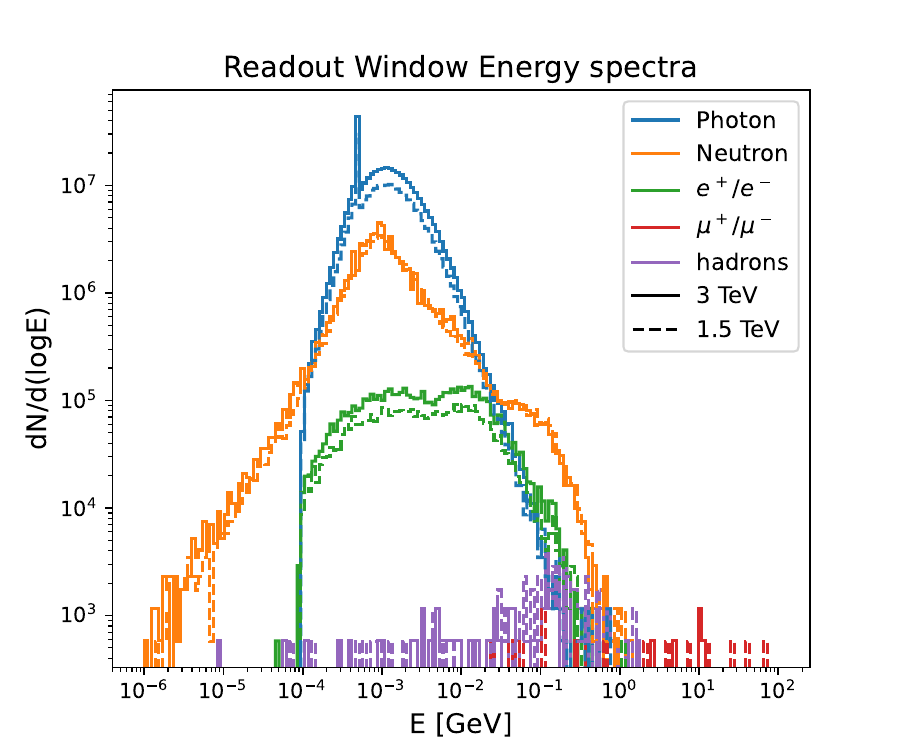}
\caption{Left: Arrival times of the main BIB components at the detector relative to the bunch crossing. The light gray band between -1 and 15 ns indicates the assumed readout window for the detector. Right: Energy spectra of the BIB particles that reach the detector within the readout window. 
The solid and dashed lines refer to the center-of-mass energy of 3 TeV and 1.5 TeV, respectively.
\label{fig:bib}}
\end{figure*}
A MuC facility consists of several sub-systems, which are described in detail in Refs.~\cite{epjc} and \cite{map-volume}. The facility structure considered for the analysis presented in this paper includes the elements discussed in Ref.~\cite{epjc}: a muon injector, an acceleration ring, and a collision ring.
Here, only the main components of a $\sqrt{s}=3$ TeV MuC complex, relevant for evaluating the background effects on the detector, are briefly outlined. 

%

After the production and acceleration of the muon and antimuon beams, they are injected into the collider ring, where single-bunch $\mu^+$ and $\mu^-$ beams circulate in the same pipe and collide head-on at the interaction point (IP). A very small bunch size is expected at the IP: $\sigma_{x,y}=3$ $\mu$m and $\sigma_{z}=5$ mm.
Due to the continuous muon decays, the beam population is exponentially decreasing. All results presented will refer to a nominal intensity of $2.2\times 10^{12}$ muons/bunch.

The Muon Accelerator Program (MAP)~\cite{MAP} in the US, designed interaction regions (IR) for muon colliders at center-of-mass energies of 125 GeV, 1.5 TeV, 3 TeV, and 6 TeV~\cite{map-acc}, and a machine-detector interface (MDI) for the 1.5 TeV collider~\cite{map-nozzle}.
The studies presented in this paper are based on MAP's configuration at $\sqrt{s}=1.5$ TeV, as a BIB sample for $\sqrt{s}=3$ TeV was not available at the start of the Higgs analyses. Later in this section, a cross-check for validating this approximation is presented.

Muons decaying in the IR are the main source of beam-induced background. Decays occurring in regions outside the IR, where dipolar fields are present, typically contribute negligibly to the BIB. Therefore, a key parameter for minimizing the background contribution is to keep the final focusing elements as short as possible. This requirement must be integrated into the lattice design process.
The IR for $\sqrt{s}=1.5$ TeV was optimized by MAP to maximize the achievable luminosity while minimizing the flux of background particles produced by the decay of muons in the beams and reaching the detector region. In fact, at the center-of-mass energy of 1.5 TeV, the dominant source of background is constituted by electromagnetic showers generated by the interactions of high-energy electrons and positrons with the machine elements. 
The solution devised by MAP to shield the detector from this irreducible background is to position two cone-shaped tungsten absorbers (referred to as ``nozzles'' throughout the paper) inside the detector along the beam pipe on both sides of the interaction point. The dimensions, shape, and materials of the nozzles have been carefully chosen to maximize the absorption of these particle fluxes while maintaining high detector acceptance.

Despite the nozzles, large fluxes of secondary and tertiary particles still reach the detector. However, their kinematical properties are such that their effects on the detector can be mitigated with proper sensor design and advanced event reconstruction methods. Figure~\ref{fig:bib} shows the main features of the particles produced in the muon decays and their subsequent interactions with the machine elements, as they enter the outer surface of the detector. This is defined as a box containing the detector, which also includes the nozzles and beam pipe boundaries. The distributions are presented for the BIB generated by MAP at $\sqrt{s}=1.5$ TeV with the MARS15 software~\cite{mars15}. 
In order to cross-check the differences between the $\sqrt{s}=1.5$ TeV and $\sqrt{s}=3$ TeV cases, a sample of beam-induced background was generated at $\sqrt{s}=3$ TeV using the MDI configuration designed for $\sqrt{s}=1.5$ TeV. This second sample was generated with FLUKA \cite{fluka}, as described in Ref.~\cite{mdi_eps}. 
Figure~\ref{fig:bib} (left) illustrates the arrival times of the background particles at the detector surface for center-of-mass energies of 1.5 and 3 TeV: a significant fraction of these are outside the assumed detector readout window of [-1, 15] ns (with $t=0$ as bunch-crossing time), and thus will not be recorded. The energy spectra of the particles entering the detector within the readout window are shown in the right panel of Fig.~\ref{fig:bib}. These particles predominantly have momenta below 1 GeV and are expected to primarily affect the detector elements closer to the background entry points. 


%
The differences in the background time and energy distributions between the two center-of-mass energies result negligible when the detector response is considered. In fact, the observed differences become negligible when the detector occupancies are calculated. This is mainly due to the different energies and detector entry angles of the particles in the two BIB samples.
The studies described in the remainder of the paper are conducted using signal and physics background samples generated at $\sqrt{s}=3$ TeV. However, as stated above, they utilize the nozzle configuration designed for $\sqrt{s}=1.5$ TeV and the BIB samples generated at the same center-of-mass energy.  

The BIB sample was generated by MAP using a detector solenoidal magnetic field of 3.57 T, which is hence employed also in the current detector model. The BIB particles, 
given at the detector entry point with their initial kinematical properties and spatial distributions, are propagated into the detector and made interact with each sub-detector by GEANT4~\cite{ref:geant4}. The detector hits produced by the BIB particle are stored on disk and then overlaid onto the hits produced by the signals and physics backgrounds to create realistic events. Further details about the BIB generation and simulation are reported in Refs.~\cite{epjc, BIBsim}.

\section{Simulation and reconstruction for signal and background samples}
\label{sec:reconstruction}

\begin{figure}[t]
    \centering
    \includegraphics[width=0.46\textwidth]{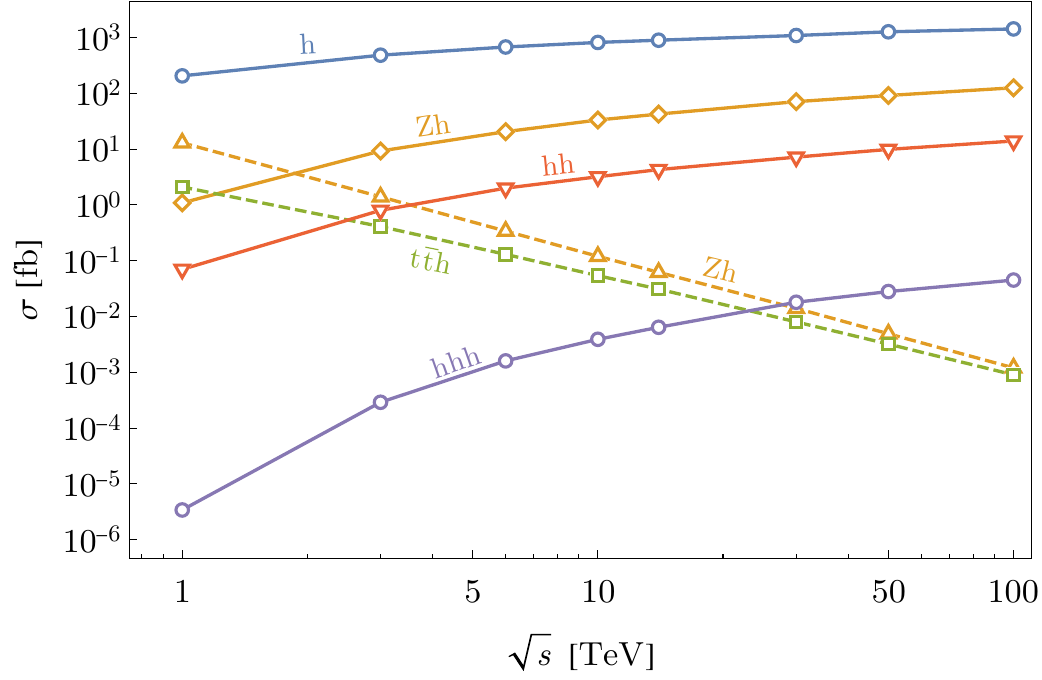}
    \caption{Cross sections for the most important Higgs boson production processes in muon-antimuon collisions as a function of the center-of-mass energy, reproduced from Ref.~\cite{musmasher}. The $ZH$ and $t\bar{t} H$ production (dashed lines) proceeds via the $s$-channel $\mu^+\mu^-$ annihilation, whereas all the other production channels are vector-boson-fusion processes. Initial-state radiation is not included in cross-section calculations.}
    \label{fig:HXS}
\end{figure}
The analysis strategy for measuring the single and double Higgs boson production cross sections is driven by their respective production mechanisms. These mechanisms determine the kinematic properties of the events as well the contributing physics background processes. 

In Sec.~\ref{sec:generators}, the generation of physics signal and background processes is described. The detector used for their detailed simulation is presented in Sec.~\ref{sec:dete}, along with the software simulation framework. Section~\ref{sec:digi} outlines the digitisation procedure, while Sec.~\ref{sec:eventreco} describes the reconstruction algorithms employed.

\begin{figure*}[t]
    \centering
    \includegraphics[width=0.3\textwidth]{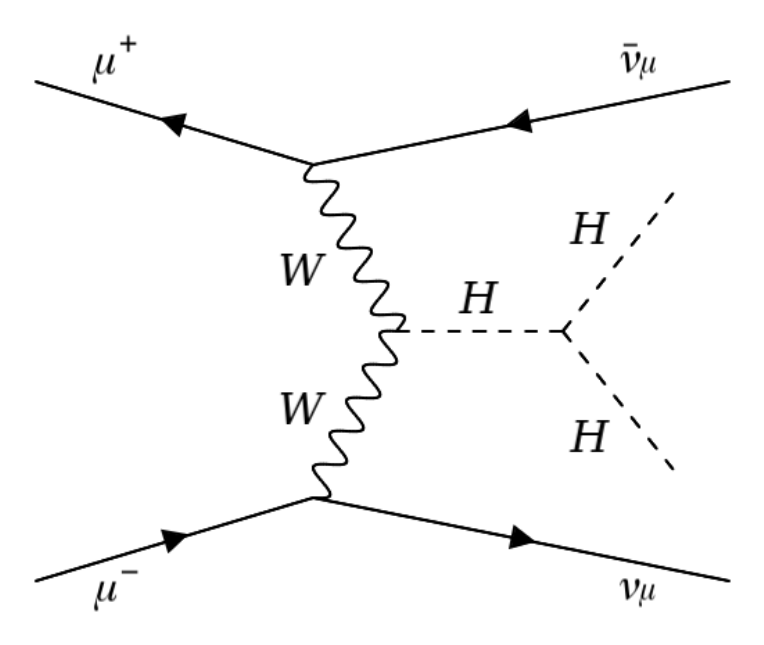} \hspace{0.4cm}
    \includegraphics[width=0.3\textwidth]{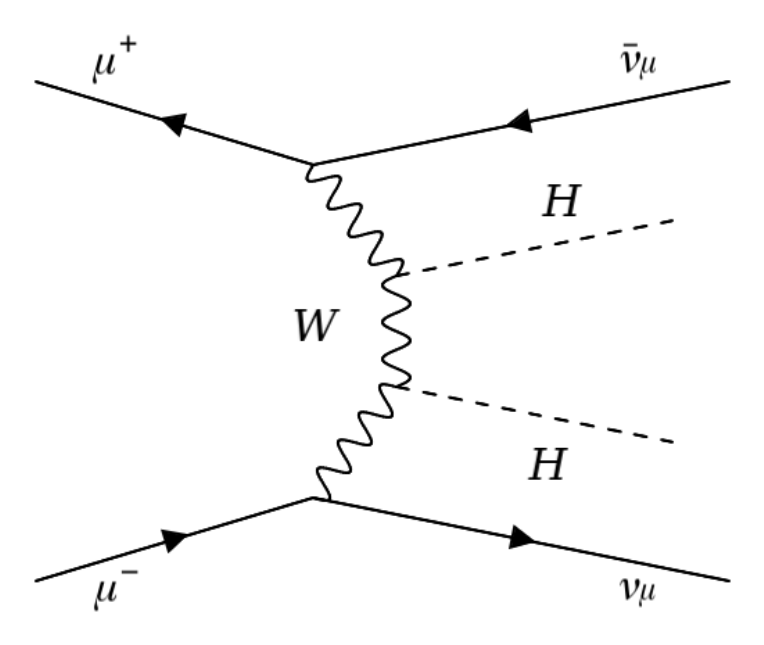} \hspace{0.4cm}
    \includegraphics[width=0.3\textwidth]{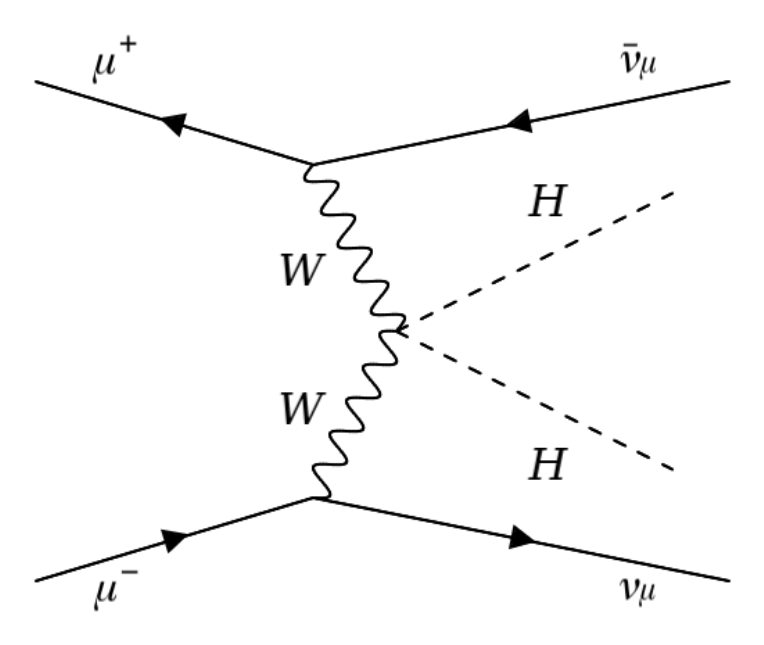}
    \caption{Feynman tree-level diagrams representing the major contributions to the double-Higgs production at a $\sqrt{s}=3$ TeV Muon Collider. The first diagram on the left represents the process directly related to the Higgs self-coupling.
\label{fig:HHFey}}
\end{figure*}
In this paper, a right-handed reference system is assumed with the origin at the center of the detector, the nominal collision point: the $z$-axis is aligned with the direction of the clockwise-circulating $\mu^+$ beam, the $y$ axis points upward, and the $x$ axis lies on the plane of the collider ring.

The MuC software framework~\cite{mucolsoft}, derived from the iLCSoft~\cite{ilcsoft} framework, has been employed for the detector simulation. iLCSoft, developed for $e^+e^-$ colliders, offers also several advanced tools for event reconstruction and analysis. Nevertheless, the MuC environment is extremely different from $e^+e^-$  colliders, due to the presence of the BIB. In most cases, the large number of BIB hits makes the use of these tools impractical due to very long computational times. In other cases, the algorithms have to be completely rethought considering the characteristics of the BIB.
For these reasons, algorithms developed specifically for the MuC are used for many tasks (like $b$-jet identification) instead of employing the state-of-the-art tools in iLCSoft.

\subsection{Signal and background physics processes}
\label{sec:generators}
\subsubsection{Higgs boson production and decay}

In muon-antimuon collisions at a center-of-mass energy of 3 TeV with unpolarized beams, the Higgs boson production is dominated by the vector boson fusion (VBF) process, as discussed in Ref.~\cite{HiggsHP} and
shown in Fig.~\ref{fig:HXS}. The $\mu^+\mu^- \to W^+W^- \nu_\mu\bar{\nu}_\mu \rightarrow H \nu_\mu\bar{\nu}_\mu$ process provides the highest contribution, followed by the $\mu^+\mu^-\rightarrow ZZ \mu^+\mu^- \rightarrow H \mu^+\mu^-$ process.

The Higgs decay modes considered in this paper are:
\begin{itemize}
    \item $H\rightarrow b\bar b$,
    \item $H\rightarrow W W^*$,
    \item $H\rightarrow Z Z^*$,
    \item $H\rightarrow \mu^+ \mu^-$,
    \item $H\rightarrow \gamma\gamma$.
\end{itemize}
The sum of their branching ratios is approximately 82\%.
Each decay mode is affected by different physics backgrounds, which are discussed in the sections dedicated to the corresponding analyses.

The double-Higgs production at $\sqrt{s}=3$ TeV is illustrated by the Feynman diagrams in Fig.~\ref{fig:HHFey}, where the dominant tree-level processes are depicted. The diagram on the left is directly related to the $\lambda_3$ parameter (see Eq.~\ref{eq:higgspotential}).
The two Higgs bosons in the final state are reconstructed in the decay mode $H\rightarrow b\bar b$, therefore the major physics background contribution comes from the process $\mu^+\mu^- \rightarrow b\bar{b}b\bar{b} \nu_{\mu} \bar{\nu}_{\mu}$.

\subsubsection{Event generation}

Signal and background samples were produced at leading order with the Monte Carlo event generators MadGraph5\_aMC@NLO~\cite{madgraph}, hereinafter referred to as Madgraph, and WHIZARD\footnote{The program version 2.8.2 was used for WHIZARD, while versions 3.X.X were used for Madgraph (differences between the versions are not relevant for this paper).}~\cite{whizard}. 
The particle hadronization and showering are performed with PYTHIA version 8.200~\cite{pythia8}.

Recently, WHIZARD authors made available a version with corrections at the next-to-leading order~\cite{whizardnlo} for multi-boson processes, but the VBF processes are not yet ready to be used. An evaluation of the impact of such corrections on the signal to noise ratio for the processes considered in this paper will be the subject of future work.

The processes generated with WHIZARD use the same setup prepared for the $e^+e^-$ studies by the CLIC Collaboration \cite{clic_higgs}. The cross sections obtained agree with previous CLIC studies up to the differences expected due to the absence of beam-strahlung in case of the MuC.

In this paper, the $Z$-boson and $W$-boson fusion processes are not distinguished experimentally, since such a separation would require the capability to detect the final-state muons of $Z$-boson fusion, which are scattered in the very forward regions of the detector. Moreover, with the requirements of Higgs analyses, the event kinematics is compatible between $Z$-boson and $W$-boson fusion productions~\cite{Castelli}.
For this reason, in most of the cases, events were generated in the $W$-boson fusion channel, which dominates the production cross sections at this center-of-mass energy. When relevant, the contribution of the $Z$-boson fusion process is evaluated by scaling the process cross section, as explained in Sec.~\ref{sec:setup}. 

Minimal requirements were applied to the sample generation, in order to speed-up the process without biasing the kinematic properties of the events, whose characteristics at this energy are not yet tested experimentally. Table~\ref{tab:MC} in Appendix \ref{app} summarises the kinematical requirements applied at generator level for each final state produced.
It should be noted that in the table and in the rest of the paper $q$ indicates any quark flavour, and in combinations like $q\bar{q}$ the quark flavours could be different (\emph{e.g.} $W \to q \bar{q}$ decay). 

\subsection{Detector simulation}
\label{sec:dete}
The detector model implemented in the detailed simulation is described in Ref.~\cite{epjc}; only the main features relevant for the reported analyses are summarised here. The MuC detector is derived from the CLIC Collaboration's detector concept~\cite{CLICREP, CLICDET}, optimized for $\sqrt{s}=3$ TeV $e^+e^-$ collisions,
and has been modified to incorporate the nozzles. 

It consists of a silicon-based tracking system surrounded by electromagnetic and hadronic calorimeters, all immersed in a 3.57 T magnetic field provided by a superconducting solenoid. The magnet iron yoke, designed to contain the return flux of the magnetic field, is instrumented with muon chambers. 

\begin{figure}[t]
\centering
\includegraphics[width=0.49\textwidth]{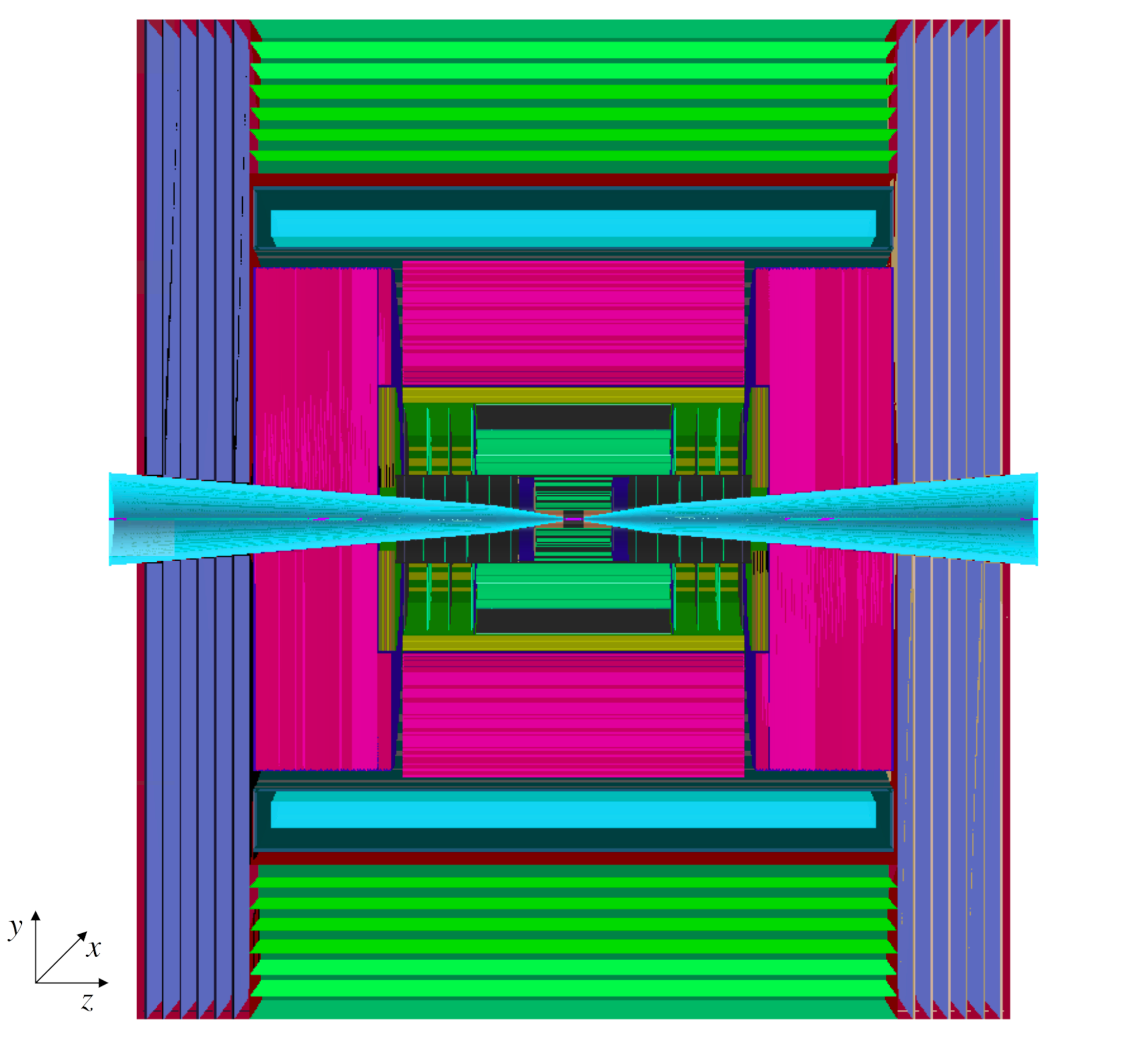}
\caption{The detector model used in the detailed simulation, view in the $y$-$z$ plane. From the innermost to the outermost regions, it includes a tracking system (green), an electromagnetic calorimeter (yellow), a hadronic calorimeter (magenta), a superconducting solenoid (light blue), barrel (light green) and endcap (blue) muon detectors. The nozzles are shown in cyan.}
\label{fig:dete}
\end{figure}

Figure~\ref{fig:dete} shows the $y$-$z$ view of the detector including the nozzles.
From the innermost to the outermost regions with respect to the IP, it features the following components:
\begin{itemize}
\item a vertex detector made of double sensor layers of $25 \times 25 ~ \mu\mathrm{m}^2$ silicon pixels, composed of four central barrel cylinders and four endcap disks on both sides of the barrel;
\item an inner tracker with three barrel layers and seven endcap disks on each side made of silicon macropixels with a size of $50 ~ \mu\mathrm{m} \times 1 ~ \mathrm{mm}$;
\item an outer tracker composed of three barrel layers and four endcap disks per side of silicon microstrips with a size of  $50 ~ \mu\mathrm{m} \times 10 ~ \mathrm{mm}$;
\item an electromagnetic calorimeter (ECAL), consisting of 40 alternating layers of tungsten absorber and silicon sensors as active material with a cell granularity of $5 \times 5$ mm$^2$, for a total of 22 radiation lengths;
\item a hadronic calorimeter (HCAL) with 60 alternating layers of steel absorber and scintillating-pad active material with a cell size of $30 \times 30$ mm$^2$, for a total of 7.5 nuclear interaction lengths;
\item a superconductive solenoid generating a magnetic field of $B=3.57$ T;
\item an iron return yoke equipped with resistive-plate chambers for muon detection, with seven layers in the barrel and six layers in the endcap, each cell having an area of $30 \times 30$ mm$^2$.
\end{itemize}
The simulation of the detector response to particle passage is performed using GEANT4, which is accessed through the MuC software framework~\cite{mucolsoft}.
%
\subsection{Digitisation}
\label{sec:digi}

A MuC event is defined as a generated signal or physics background process with the BIB particles produced in a bunch crossing overlaid.  
The merging is performed at the level of detector hits, ensuring that both contributions, the signal or the physics background and the BIB,  undergo hit digitization in each sensor of the sub-detectors simultaneously. This simulates the process as it would occur during data collection.
Marlin~\cite{marlin}, one of the packages that the MuC software framework adopted from iLCSoft, performs these actions.

A parameterized digitisation is used for the tracker: hit positions and times are smeared with Gaussian distributions representing the spatial and timing resolutions.
For ECAL, HCAL, and the Muon System, a digitization is applied where the energies of hits in the same cell are summed up, and the time of the earliest hit is assigned as the time of the digitized hit.

Only hits with arrival times within a specified range relative to the bunch-crossing time ($t=0$) are retained for each sensor to reduce the impact of the BIB component. These time windows are applied on the normalized hit time, defined as $t_N = t - D/c$, where $t$ is the absolute time of the hit, $D$ is the distance of the sensor from the interaction point, and $c$ is the speed of light. The time windows are: $[-90,150]$ ps for the vertex detector and $[-180,300]$ ps for the other tracking systems, while for ECAL, HCAL and Muon System hits the range is $\pm 250$ ps.
To meet these time requirements, an excellent time resolution is essential. A resolution of 30 (60) ps for the vertex (tracking) system and 100 ps for the calorimeters is assumed in the detailed detector simulation. These time resolutions are expected to be achievable with the technologies developed for HL-LHC, as described in Ref.~\cite{epjc}. Cooling system and cables are not included in the simulation, but it is expected that the impact of service structures will be minimized with the developments over the next 20 years.

\subsection{Event reconstruction}
\label{sec:eventreco}

The reconstruction algorithms of the physics objects, inherited from the iLCSoft package~\cite{ilcsoft}, have been adapted for the MuC background conditions, although a complete optimization is still in progress.
The Marlin software is employed for the reconstruction of the physics objects. Marlin also makes use of other packages for specific reconstruction tasks, such as the PandoraPFA package~\cite{pandora}, which is used for particle flow techniques.
A brief description of the utilized algorithms is provided below:
\begin{itemize}
\item \textbf{Tracks}: The trajectories of charged particles (tracks) are reconstructed from the positions of the hits in the silicon tracking stations. Hits are clustered into tracks using two different algorithms. The first one is the Conformal Tracking (CT)~\cite{CT}, while the second one uses the Combinatorial Kalman Filter, henceforth referred to as the CKF algorithm~\cite{CKF}. The CT algorithm was originally designed for the clean environment of electron-positron collisions and has long running times in the MuC environment due to the presence of the BIB hits (order of hours per event per CPU core with the available computational resources). For this reason, it is usually applied to pre-selected hits in regions of interests, \emph{e.g.} inside calorimetric jet cones or around reconstructed segments of the muon detectors. 
The MuC framework uses a version of the CKF algorithm that was tuned for LHC proton-proton collisions and runs in the MuC environment with much shorter execution times (order of minutes per event). 
%
\begin{figure}[t]
    \centering
    \includegraphics[width=0.45\textwidth,trim={2mm 2mm 2mm 2mm},clip]{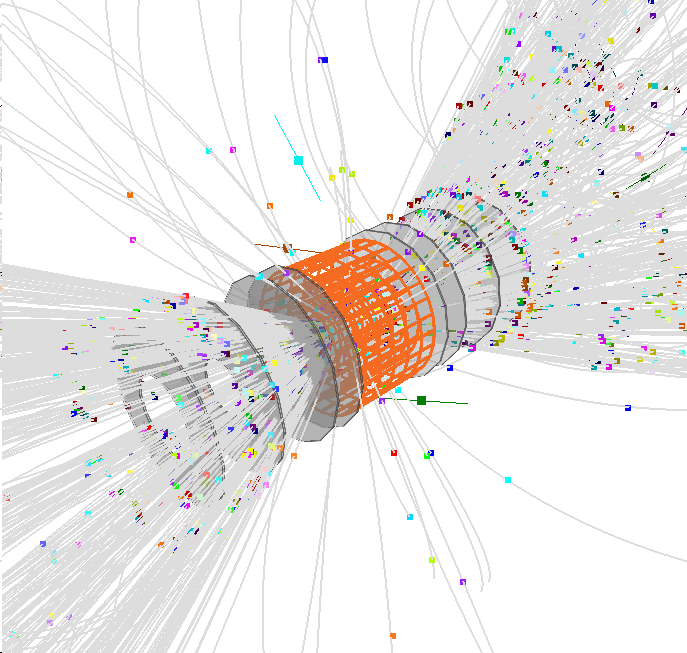}
    \caption{An image of the vertex detector with the barrel sensor pads in orange and the endcap disks in dark gray. The colored squares represent the hits produced by the BIB in a single bunch crossing, while the light gray lines show the reconstructed fake tracks. It can be observed that the majority of the tracks are reconstructed in the forward and backward regions, where the density of BIB hits is higher.
    \label{fig:eveDisplay}}
\end{figure}
An example of the fake tracks reconstructed from the BIB hits with the CKF algorithm in the central region of the tracking system for a single bunch crossing is shown in Fig.~\ref{fig:eveDisplay}.
The uncertainty on the transverse momentum, $\Delta p_{\mathrm{T}}$, obtained with the CT and CKF algorithms goes from $\Delta p_{\mathrm{T}} / p^2_{\mathrm{T}} \approx 1 \cdot 10^{-1}$ $(5 \cdot 10^{-3})$ GeV$^{-1}$ for muons with momentum $p=1$ (100) GeV and polar angle $\theta = 13^{\circ}$, to $\Delta p_{\mathrm{T}} / p^2_{\mathrm{T}} \approx 5 \cdot 10^{-2}$ $(4 \cdot 10^{-5})$ GeV$^{-1}$ for muons with $p=1$ (100) GeV and $\theta = 89^{\circ}$. The track reconstruction efficiency with the CT algorithm is approximately 75\% at $\theta=13^{\circ}$ and close to 100\% at $\theta=89^{\circ}$. 
In this paper, the CT algorithm is used for the $H \to WW^*$ and $H \to \mu^+ \mu^-$ analyses, whereas the CKF algorithm is employed for the $H \to b \bar{b}$, $H \to ZZ^*$, $HH \to b \bar{b}b \bar{b}$, and Higgs width analyses.
\item \textbf{Photons}: Photons are defined as clusters of ECAL hits that are not matched to tracks.
For photon reconstruction, the PandoraPFA algorithm is used, which takes tracks and calorimeter hits as inputs. For the purpose of photon identification, tracks are reconstructed using the CKF algorithm and track quality requirements are applied~\cite{epjc}. An energy threshold of 2 MeV is applied to the calorimeter hits to reduce the BIB contamination.
This threshold has a significant impact on the reconstruction of the jets, as will be discussed later. 
%
The photon reconstruction efficiency is about 78\% for energies below 50 GeV and about 98\% for energies above 300 GeV. The photon energy resolution is approximately 13\% for energies below 50 GeV and 1.5\% for energies above 400 GeV.
\item \textbf{Muons}: The PandoraPFA algorithm is also used to reconstruct and identify muons: tracks reconstructed with either the CT or CKF algorithm are matched to stubs in the muon system. 
The requirements on the angle of the muon tracks suppress the BIB contamination, that is higher in the forward region of the muon system, at the cost of a low reconstruction and identification efficiency in the forward and backward regions. 
The efficiency, evaluated in this paper for the CKF tracking algorithm, is about 20$\%$ in the angular regions $10^{\circ} < \theta < 20^{\circ}$ and $160^{\circ} < \theta < 170^{\circ}$, while it is close to 100$\%$ for $40^{\circ} <\theta<140^{\circ}$. The uncertainty on the muon transverse momentum goes from $\Delta p_{\mathrm{T}} / p^2_{\mathrm{T}} \approx 2.1 \cdot 10^{-3}$ GeV$^{-1}$ for muons with $p_{\mathrm{T}}=2.5$ GeV to $\Delta p_{\mathrm{T}} / p^2_{\mathrm{T}} \approx 1 \cdot 10^{-4}$ GeV$^{-1}$ for muons with $p_{\mathrm{T}}>25$ GeV. The choice of the tracking algorithm employed in a specific Higgs analysis, CT or CKF, depends mainly on the available computing resources for each study.
\item \textbf{Jets}: Hadronic jets are reconstructed with the following procedure: 
\begin{enumerate}
\item tracks are reconstructed as described above and filtered to remove combinatorial background by requiring a minimum number of hits on different tracking layers~\cite{epjc};
\item calorimeter hits are selected requiring an energy threshold of 2 MeV to remove spurious BIB hits, as in photon reconstruction; 
\item tracks and calorimeter hits are used as inputs to the PandoraPFA algorithm to obtain the reconstructed physics objects;
\item the $k_t$ jet algorithm~\cite{kt} with a radius parameter of 0.5 is used to cluster the reconstructed physics objects into jets.
\end{enumerate}
This reconstruction procedure has a jet finding efficiency of $85\%$ for a jet $p_{\mathrm{T}}$ around 20 GeV, and $90\%$ for a jet $p_{\mathrm{T}}$ around 200 GeV. On average, with a $p_{\mathrm{T}}$ threshold of 10 GeV, 13 fake jets are reconstructed per event due to the BIB contribution. Higgs analyses involving jets in the final state apply cleaning requirements to suppress the fake rate.
More information about fake jets and their mitigation can be found in Ref.~\cite{epjc}.

The jet 4-momentum is calculated by summing the 4-momenta of the clustered physics objects, and the jet direction is defined by the jet momentum direction. A correction is applied to the reconstructed jet energy to take into account detector effects and inefficiencies. The correction is calculated as a function of jet $p_{\mathrm{T}}$ and $\theta$ by comparing reconstructed jets with corresponding truth-level jets (\emph{i.e.} jets clustered starting from generator-level stable and detectable particles).
The jet $p_{\mathrm{T}}$ resolution is $35\%$ for a jet $p_{\mathrm{T}}$ around 20 GeV, and above $20\%$ for a jet $p_{\mathrm{T}}$ around 200 GeV.

\item \textbf{$\bm{b}$-tagging}: The identification of jets originating from $b$ quarks and their separation from $c$- and light-quark jets is performed by searching for displaced secondary vertices (SV) within the jets~\cite{DaMolin}. The CT algorithm is used to reconstruct tracks inside the jet cones. A jet is identified as a $b$ jet if a secondary vertex is found with a proper lifetime $\tau=L\,m/P$ greater than 0.2 ps, where $L$ is the distance between the primary and the secondary vertices, $m$ is the mass of the SV, and $P$ is the momentum associated to the SV, calculated as the sum of the momenta of the tracks associated to it.
The $b$-jet identification efficiency and misidentification rate have been determined as a function of the jet $p_{\mathrm{T}}$ and $\theta$ with independent samples of $b\bar{b}$, $c\bar{c}$ and light-quark jets~\cite{epjc}.
The tagging algorithm has an efficiency of about $45\%$ for a $b$-jet $p_{\mathrm{T}}$ around 20 GeV, and of about $70\%$ for a $b$-jet $p_{\mathrm{T}}$ around 140 GeV with a misidentification rate of about $20\%$ for $c$-jets and from $0.1\%$ to $5\%$ for light-quark jets. 
It should be noted that the fake rate in the case of tagged jets is negligible.
The algorithm used in the analyses presented in this paper is very simple, a much more advanced algorithm based on artificial-intelligence methods is planned for future updates.
%
%
\end{itemize}
A detailed discussion of the reconstruction performance of these objects can be found in Ref.~\cite{epjc}.
%

\subsection{
Analysis path}
\label{sec:setup}


\begin{table*}[h]
    \centering
    \caption{Theoretical cross section in the kinematical fiducial region, pre-selection efficiency, $b$-tagging efficiency, and expected events for the signal and background processes in the $H \to b \bar{b}$ analysis for $\mu^+\mu^-$ collisions at 3 TeV and $\mathcal{L}=1$ ab$^{-1}$. $X$ indicates both $\nu\bar{\nu}$ and $\ell\ell$.}
    \begin{tabular}{c|c c c c}
    \hline
    \hline
        \textbf{Process} &  $\sigma \ [\mathrm{fb}]$& $\epsilon_{presel} [\%]$ & $\epsilon_{tag} \ [\%]$ & $N_{exp}$ \\
    \hline
       $\mu^+ \mu^- \to H (\to b \bar{b}) X $   & $308$  & $39.2$  & $49.3$   & $59520$ \\
       \hline
       $\mu^+ \mu^- \to H (\to c \bar{c}) X $   & $15.3$ & $39.2$  & $29.6$   & $1775$  \\
       $\mu^+ \mu^- \to H (\to g g) X $         & $43.0$ & $39.2$  & $0.0021$ & $0.35$  \\
       $\mu^+ \mu^- \to H (\to WW^* \to q\bar{q}q\bar{q}) X $         & $50.8$ & $64.0$  & $6.88$ & $2237$  \\
       $\mu^+ \mu^- \to  q\bar{q} X$ & $2600$ & $23.5$  & $10.7$  & $65430$ \\
       $\mu^+ \mu^- \to  q\bar{q} \ell \nu$     & $15000$ & $5.74$   &$0.14$   & $1200$   \\ 
      \hline
      \hline
    \end{tabular}
    \label{tab:hbb}
\end{table*}
%
The analyses presented below focus exclusively on the dominant $WW$- and $ZZ$-fusion production processes (WWF and ZZF) of the Higgs boson, which are referred to with the compact notation $\mu^+ \mu^- \rightarrow H X$, where $X=\nu \nu$ or $\ell \ell$, that stands for a couple of neutrinos or in alternative for a couple of leptons. 
Since their kinematic properties are found to be compatible at $\sqrt{s}=3$ TeV~\cite{Castelli}, only the WWF process has been generated for the signal samples, the kinematic distributions of which are considered representative of the entire signal.
The total signal cross section is then obtained by scaling the WWF cross section by the ratio of the WWF to ZZF cross sections ($f$):
\begin{multline}
\label{eq:wzscale}
\sigma_H = \sigma(\mu^+ \mu^- \rightarrow H X) = \\ \sigma(\mu^+ \mu^- \rightarrow H \nu \bar{\nu}) \cdot \left(1+\frac{1}{f}\right).
\end{multline}
The factor $(1+1/f)$ is found to be 1.09 using both WHIZARD and Madgraph.
Exceptions to this procedure will be explained in the text.

The integrated luminosity is assumed to be $\mathcal{L}=1$ ab$^{-1}$, which is expected to be collected in 5 years of data-taking by a single experiment \cite{epjc}. The possibility of a longer run of 10 years that would provide $\mathcal{L}=2$ ab$^{-1}$ has been considered in other publications and presentations. 

The backgrounds considered in each analysis and reported in the corresponding tables, are those that give a significant contribution to the expected number of selected events, and they have been identified depending on their cross sections and selection efficiencies.
Backgrounds without resonances are produced via WWF and ZZF processes, or other sub-dominant diagrams that have the same final-state particles. In these cases, two neutrinos or two leptons that are usually very forward and not detected are present. In a similar approach to the Higgs processes, in some analyses these are indicated with the compact notation $X=(\nu\bar{\nu},\ell\ell)$.
The signal and background selection efficiencies ($\epsilon$) are defined as the ratio between the number of events passing the selection requirements and the corresponding number of generated events. The number $N_{\mathit{exp}}$ of expected signal and background events is determined with the formula:
\begin{equation}
    \label{eq:nexp}
N_{\mathit{exp}} = \epsilon \cdot \sigma \cdot \cal{L}\ ,
\end{equation}
where $\sigma$ is the production cross section given by the event generator.

The following studies on the Higgs boson cross sections estimate the statistical sensitivity on the production cross section times the branching fraction $\sigma_H \times BF(H \to xx)$, where $xx$ denotes any considered final state. For the sake of compactness, the quantity $\sigma_H \times BF(H \to xx)$ is indicated with $\sigma(H \to xx)$, and the corresponding uncertainty with $\Delta \sigma(H \to xx)$.
It is important to note that in the studies only the statistical sensitivity, resulting from the number of estimated events in the selected samples, is evaluated. A discussion of systematic uncertainties is presented in Sec. \ref{sec:sys}.

In several Higgs analyses, machine-learning methods such as Boosted Decision Trees (BDTs)~\cite{bdt} and Multi Layer Perceptrons (MLPs) are employed to enhance the separation between the signal and the physics backgrounds. The BDTs and MLPs have been trained and applied with the TMVA package~\cite{hoecker2009tmva} from the ROOT software suite~\cite{Brun:1997pa}. 
Their configuration parameters are set to the default values as defined by TMVA.

\section{Determination of Higgs boson cross sections times branching ratios}
\label{sec:higgsxs}

\subsection[$H\to b \bar{b}$ cross section]{$H\to b \bar{b}$ cross section}
\label{sec:h2bb}

This section is dedicated to the determination of the statistical sensitivity on the measurement of the $H \rightarrow b \bar{b}$ cross section.
Samples of $\mu^+ \mu^- \rightarrow  H(\rightarrow b \bar{b})X$ signal events and background events have been generated following the prescription of Sec. \ref{sec:generators}, and are reported in Tab. \ref{tab:hbb}.

Jets are reconstructed as explained in Sec.~\ref{sec:eventreco} using the CKF algorithm for the input tracks. Jets with transverse momentum $p_{\mathrm{T}}>40$ GeV and pseudorapidity\footnote{The pseudorapidity is defined as $\eta = - \mathrm{log} \left ( \mathrm{tan} \frac{\theta}{2} \right)$, where $\theta$ is the angle of the particle momentum with respect to the beam axis.} $|\eta|<2.5$ are selected to maximize the reconstruction performance and minimize the BIB effects.
Both jets from the Higgs decay are required to be identified as $b$-jets, \emph{i.e.} a displaced SV has to be associated to them. 
The identification efficiencies and mis-identification rates have been determined as a function of jet $p_{\mathrm{T}}$ and $\theta$ (Sec.~\ref{sec:eventreco}) and applied to the signal and background samples as event weights. The mis-identification rate of light-quark jets has been found to be negligible, and for this reason light-quark jets have not been included in the background processes. The fake jet rate is negligible after the SV requirement.

\begin{figure}[b]
\centering
\includegraphics[width=0.48\textwidth]{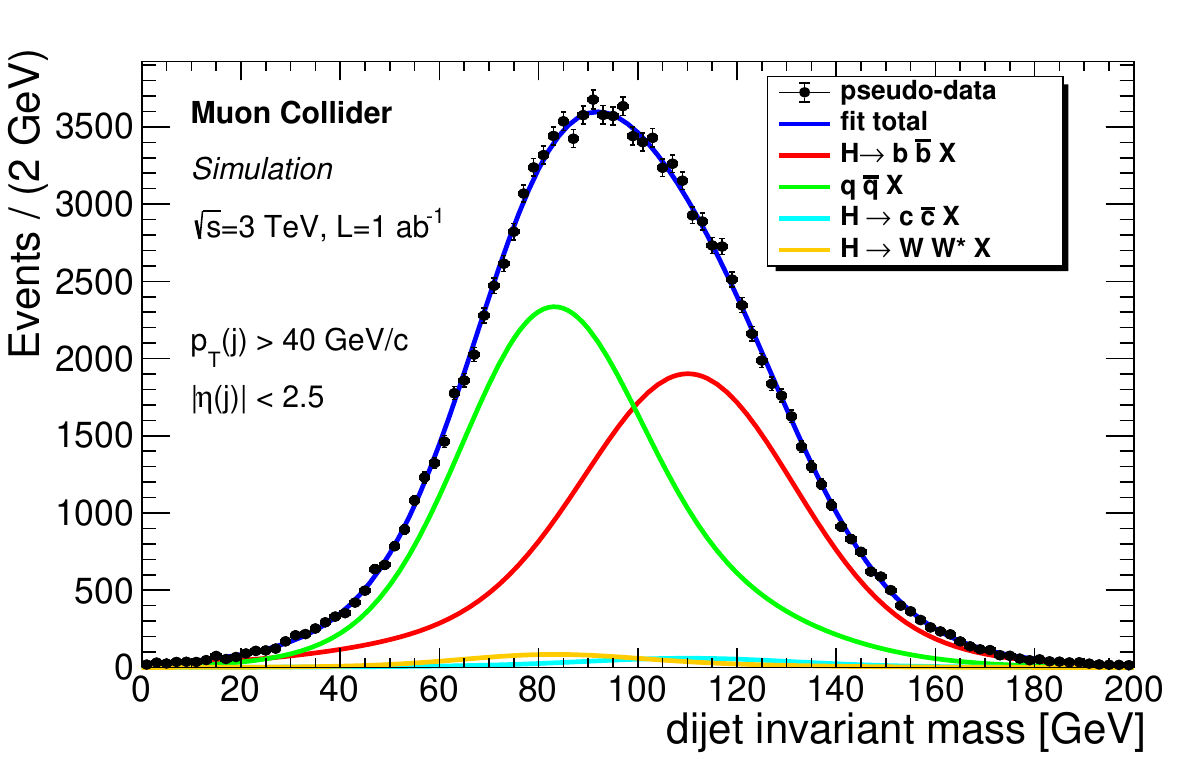}
\caption{Example of the di-jet invariant mass fit for one of the pseudo-experiments used to extract the $H \rightarrow b \bar{b}$ yield and uncertainty. 
    Pseudo-data are obtained assuming $\mu^+\mu^-$ collisions at 3 TeV and $\mathcal{L}=1$ ab$^{-1}$ \cite{annual}. In the legend, $X$ stands for both $\nu\bar{\nu}$ and $\ell\ell$. For this plot the detailed simulation of the detector has been used, but $b$-tagging efficiencies are taken as event weights from previous studies with BIB, as explained in the text.
\label{fig:hbb_fit}}
\end{figure}

The expected number of signal and background events is determined with Eq.~\ref{eq:nexp}: about 60k $H \rightarrow b \bar{b}$ events and 70k background events are estimated to be collected in 1 ab$^{-1}$, as reported in Tab.~\ref{tab:hbb}.
The di-jet invariant-mass distributions for signal and background events are fitted with double-Gaussian functions to obtain signal and background models
that are used to build a likelihood function and generate samples of pseudo-data according to the expected number of events.
An unbinned maximum-likelihood fit is then performed on the invariant-mass distributions of the pseudo-experiments, allowing the signal and the main background ($q\bar{q} \nu\bar{\nu}$) yields to float. The contributions from $H \to c \bar{c}$ and $H \to WW^*$ are fixed in the fit.
In this way, the average $H \rightarrow b \bar{b}$ yield and its uncertainty are extracted. 
An example of one of the fits is shown in Fig.~\ref{fig:hbb_fit}. Due to the limited jet energy resolution, the di-jet invariant-mass distribution shows a significant overlap between the Higgs signal and the background, consisting mainly of $Z$ boson decays to $b\bar{b}$ and $c\bar{c}$ pairs. In any case, the number of events is sufficiently high to disentangle the two components on a statistical basis.
The $H/Z$ separation will be improved by utilizing a jet reconstruction algorithm that is specifically tuned and optimized for MuC.
A systematic uncertainty on the fact that $H \to c \bar{c}$ and $H \to WW^*$ are fixed in the fit should be considered. However, assuming a very conservative 20\% uncertainty on the prediction of their cross sections, an uncertainty on the $H \to b \bar{b}$ yield in the order of 0.5\% is obtained. A discussion on other sources of systematic uncertainty is presented in Sec. \ref{sec:sys}.

\begin{figure*}[h]
\centering
\includegraphics[width=0.45\textwidth]{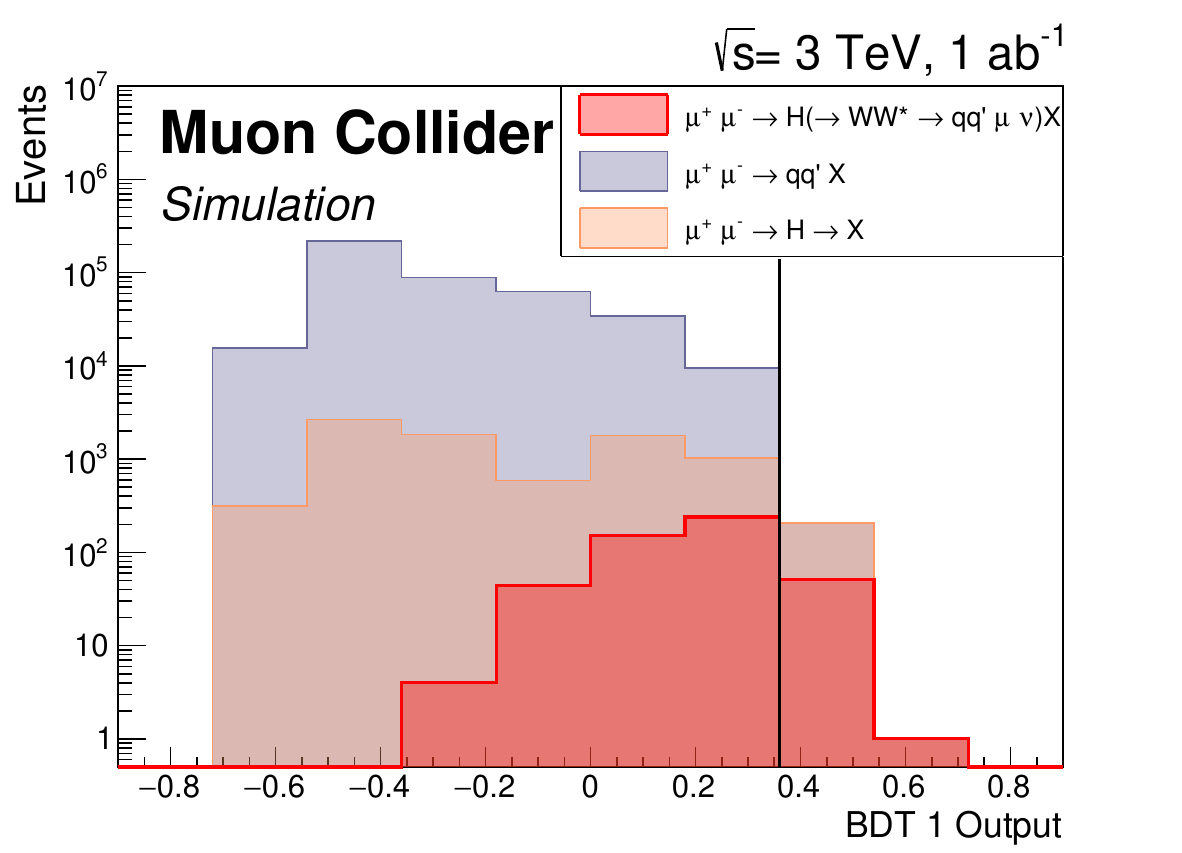}
\includegraphics[width=0.45\textwidth]{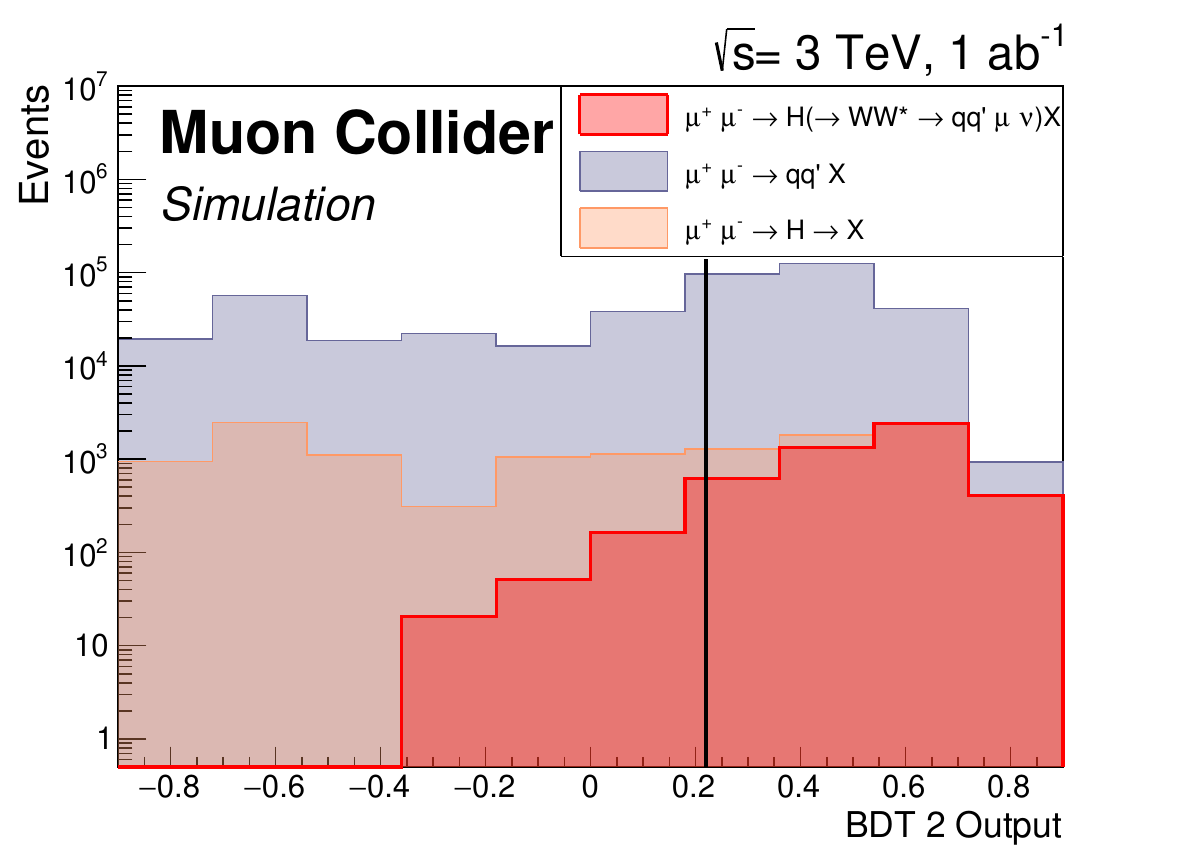}
\caption{Distributions of the BDT outputs to separate the $H \to WW^*$ signal from the background processes without a Higgs bosons (left) and with a Higgs boson (right). Signal and backgrounds are presented with stacked histograms, normalized to the expected number of events. The black vertical lines indicate the threshold for achieving the optimal signal significance. In the legend, $X$ stands for both $\nu\bar{\nu}$ and $\ell\ell$.}
\label{fig:hwwbdt}
\end{figure*}
The cross section is estimated from Eq.~\ref{eq:nexp}, using the expected number of signal events and the total signal selection efficiency.
Assuming the uncertainties on the efficiency and the luminosity negligible, the statistical uncertainty on the cross section is dominated by the uncertainty on the signal yield. By averaging the results of several pseudo-experiments, the estimated uncertainty on the production cross section is:
\begin{equation}
    \frac{\Delta \sigma (H\to b\bar{b})}{\sigma (H\to b\bar{b})} = 0.78\%\ .
\end{equation}
%
%
\subsection[$H \rightarrow WW^*$ cross section]{$H \rightarrow WW^*$ cross section}
\label{sec:h2ww}

The $H\to WW^*$ decay is reconstructed in the semileptonic final state, where one $W$ boson decays into a muon and a muonic neutrino and the other one decays hadronically into two jets: $H\to WW^* \to q\bar{q}\mu \nu$~\cite{Castelli}. This final state has been chosen for its good signal-over-background ratio. In fact, the selection of a high-momentum muon significantly reduces the background from full hadronic final states. On the other hand, the $W$ boson decay to jets has a high branching ratio, hence the semileptonic final state provides higher statistics than the fully leptonic final state.

Signal and background samples have been generated as summarized in Tab.~\ref{tab:MC}. They have been grouped into two sets, depending on the presence of the Higgs boson in the process: 
\begin{enumerate}
    \item background processes without a Higgs boson: $\mu^+ \mu^- \to q\bar{q} \ell \nu$, $\mu^+\mu^- \to q\bar{q}\ell\ell$, and $\mu^+\mu^- \to q\bar{q}\nu\bar{\nu}$, with $q = u, d, c, s$ and $\ell = e, \mu, \tau$; 
    \item background processes with a Higgs boson: $\mu^+\mu^- \to H (\to WW^* \to q\bar{q}q\bar{q})X$, $\mu^+\mu^- \to H (\to b \bar{b})X$, and $\mu^+\mu^- \to H (\to \tau^+ \tau^-) X$.
\end{enumerate}
In this analysis, jets are first clustered using calorimeter hits, then tracks are reconstructed with the CT algorithm in regions of interest defined by the calorimetric jet cones. Tracks obtained in this way, together with calorimeter clusters, are given as input to the final jet reconstruction. Muons are reconstructed using CT tracks as input. In order to speed-up the muon reconstruction, BIB hits are neglected for this specific task: this approximation does not bias the result since the muon reconstruction requirements suppress the BIB contribution, at the cost of a low efficiency in the forward and backward regions (Sec.~\ref{sec:eventreco}). 
%
%
Signal candidates are formed by combining two reconstructed jets and one muon.
The muon is required to have $p_{\mathrm{T}} > 10$ GeV and $|\eta| < 2.44$ (corresponding to $10^\circ < \theta < 170^\circ$). The two jets must be central with $|\eta| < 2.5$ and $p_{\mathrm{T}}> 20$ GeV.
Requirements are applied to remove fake jets from the analysis: the number of the jet constituents must be greater than two, and the maximum fraction of the jet momentum carried by a single constituent is required to be less than 0.8. Further details on this prescription can be found in Ref.~\cite{Castelli}.

\begin{table*}[h!]
    \centering
    \caption{Theoretical cross sections within the kinematical fiducial region, pre-selection efficiency, BDT efficiency, and expected number of events for signal and background processes in the $H \to WW^{*}$ analysis with $\mu^+\mu^-$ collisions at 3 TeV and $\mathcal{L}=1$ ab$^{-1}$. $X$ indicates both $\nu\bar{\nu}$ and $\ell\ell$. The upper limits are determined at $68 \%$ Confidence Level.}
    \begin{tabular}{c|c c c c}
    \hline
    \hline
         \textbf{Process} &  $\sigma \ [\mathrm{fb}]$ & $\epsilon_{presel} [\%]$ & $\epsilon_{BDT} [\%]$ & $N_{exp}$ \\
    \hline
       $\mu^+ \mu^- \to H (\to W W^* \to q\bar{q} \mu \nu) X$ & $17.3$ & $ 29.0$ & $ 48.5$ & $2430$\\
       \hline
       $\mu^+ \mu^- \to H (\to W W^* \to q\bar{q}q\bar{q})X $ & $ 50.8$ & $ 3.9$ & $ <0.2$ & $<5$\\
       $\mu^+ \mu^- \to H ( \to b\bar{b})X$ & $ 308 $ & $ 1.1$ & $ <4.4$ & $<150$\\
       $\mu^+ \mu^- \to H( \to \tau^+ \tau^-)X$ & $34.3$ & $0.3 $ & $<4.0$ & $<4$ \\
        $\mu^+ \mu^- \to  q\bar{q} \ell\nu $ & $15000 $ & $2.1$ &$0.81$ & 2600 \\
        $\mu^+ \mu^- \to  q\bar{q} \ell\ell$ & 1040 & $2.3$ &$ <0.42$ &  $<100$ \\
        $\mu^+ \mu^- \to  q\bar{q} \nu\bar{\nu} $ & 1560 & $4.6$&$ <0.14$ & $<100$ \\   
      \hline
      \hline
    \end{tabular}
    \label{tab:hww}
\end{table*}
In order to separate signal from backgrounds, two BDTs have been trained: the first to separate the $H \to WW^*$ signal from the background processes without a Higgs bosons (listed above in bullet 1), the second to separate the signal from the processes with a Higgs boson (bullet 2).
The BDTs have the following observables as inputs:
\begin{itemize}
\item the $p_{T}$ and $\theta$ of the selected muon;
\item the invariant mass, $p_{T}$, and $\theta$ of the $W \to q\bar{q}$ candidate formed by two jets;
\item the invariant mass, $p_{T}$, and $\theta$ of the Higgs boson candidate;
\item the missing transverse momentum, defined as $P^{\mathrm{miss}}_{T} = - \sqrt{P_x^2+P_y^2}$ with $P_{x,y} = \sum_{i}p_{x,y}^{\mu_i} + \sum_{i}p_{x,y}^{\mathrm{jet}_i}$,
where the momentum components $p_x$ and $p_y$ are summed over all reconstructed muons and jets in the event;
\item the number of isolated muons in the event, \emph{i.e.} the muons with $\Delta R=\sqrt{\Delta\phi^2+\Delta\eta^2}>0.5$ with respect to any jet in the event;
\item three acollinearity angles, defined between the momenta $\vec{p}_i$ and $\vec{p}_j$ of two particles as:
\begin{equation}
\label{eq:acol}
        \theta_{\mathrm{acoll}}(i,j) = \pi - \mathrm{acos}\left(\frac{\vec{p}_i \cdot \vec{p}_j}{p_i p_j} \right)\ ,
\end{equation}
where $p_{i,j} = |\vec{p}_{i,j}|$ and $i$ and $j$ are: the Higgs boson candidate, the $W \to q\bar{q}$ candidate, and the selected muon. Hence, the three acollinearity angles are $\theta_{\mathrm{acoll}}(H,\mu)$, $\theta_{\mathrm{acoll}}(H,W \to q\bar{q})$, and $\theta_{\mathrm{acoll}}(W \to q\bar{q},\mu)$.
\end{itemize}
\begin{table*}[t]
    \centering
    \caption{Theoretical cross section within the kinematical fiducial region, pre-selection efficiency, BDT efficiency, and expected events for signal and background processes in the $H \to ZZ^{*}$ analysis with $\mu^+\mu^-$ collisions at 3 TeV and $\mathcal{L}=1$ ab$^{-1}$.}
    \begin{tabular}{c|c c c c}
    \hline
    \hline
         \textbf{Process} & $\sigma \ [\mathrm{fb}]$ & $\epsilon_{presel} [\%]$ & $\epsilon_{BDT} [\%]$ & $N_{exp}$ \\
    \hline
       $\mu^+\mu^- \to H(\to Z Z^*) X \to q\bar{q} \mu^+\mu^- $ & $0.376$ & 25.4 & 63.8 & $60$ \\
       \hline
       $\mu^+\mu^- \to q\bar{q} \mu^+\mu^- (\nu \bar{\nu}, \ell \ell)$ & $6.18$ & 15.0  & 4.53 & $42$ \\
      \hline
      \hline
    \end{tabular}
    \label{tab:hzz}
\end{table*}
More details on the BDT input variables can be found in Ref.~\cite{Castelli}.
The output of the BDTs for the signal and backgrounds is shown in Fig.~\ref{fig:hwwbdt}.
The outputs of the two BDTs are required to be above thresholds determined by maximizing the signal significance 
\begin{equation}
\label{eq:significance}
\mathcal{S}=\frac{S}{\sqrt{S+B}}\ ,
\end{equation}
where $S$ and $B$ are the expected number of signal and background events, respectively, calculated using Eq.~\ref{eq:nexp}.
Table~\ref{tab:hww} summarises the theoretical cross sections, the efficiencies, and the expected number of events passing the described selection.  
After the event selection, the dominant physics background is the process $\mu^+ \mu^- \to  q\bar{q} \mu\nu$.

The statistical sensitivity on the cross section is determined considering this case as a counting experiment. 
Thus, assuming that the uncertainties on the efficiency and the integrated luminosity are negligible, the error on the $H \to WW^*$ cross section is given by:
\begin{equation}
    \frac{\Delta \sigma (H\to WW^*)}{\sigma (H\to WW^*)} = \frac{\sqrt{S+B}}{S}\ ,
\end{equation}
where the formula is derived considering the properties of the Poisson statistics.
The statistical sensitivity resulting from the numbers in Tab.~\ref{tab:hww} is:
\begin{equation}
    \frac{\Delta \sigma (H\to WW^*)}{\sigma (H\to WW^*)} = 2.9\%\ .
\end{equation}
%
%
\subsection[$H \rightarrow ZZ^{*}$ cross section]{$H \rightarrow ZZ^{*}$ cross section}
\label{sec:h2zz}

As with the $H \to WW^*$ channel, the $H \to ZZ^*$ production has been studied in the semileptonic final state, where one $Z$ boson decays into two muons and the other into two jets: $\mu^+\mu^- \to H(\to Z Z^*) X \to q\bar{q} \mu^+\mu^- X$. 
Differently from the $H \to WW^*$ analysis, the $Z \to \mu \mu$ decay is fully reconstructed, 
and the selection requirements described below reduce the background to a negligible level.
On the other hand, the branching ratio of $H \to ZZ^*$ is approximately 10 times smaller than that of $H \to WW^*$, and in this channel the irreducible background $\mu^+\mu^- \to q\bar{q} \mu^+\mu^- X$ becomes dominant. 

The input tracks of jets and muons have been reconstructed by employing the CKF tracking algorithm. To minimize the number of fake jets, at least one track is required in the jets. 
Jets must have transverse momentum $p_{\mathrm{T}}>$15 GeV and $|\eta|<1.32$ (corresponding to $30^\circ < \theta < 150^\circ$). Muons are selected with $p_{\mathrm{T}}>10$ GeV. It should be noticed that the angular region selected is smaller than that used for the $H \to WW^*$ analysis, since it has been preferred to consider the central region of the calorimeters, with very low contamination from fake jets, which are concentrated in the forward region.
A signal candidate is formed by the combination of two jets and two opposite-charge muons.
In order to remove background processes where muons are produced in the jet fragmentation, the distance $\Delta R$ in the $\eta$-$\phi$ space between muons and jets for each combination must be greater than 0.5.  
If more than one combination per event is found, the candidate with the invariant mass closer to the Higgs boson mass is selected.

To remove the irreducible background $\mu^+\mu^- \to q\bar{q} \mu^+\mu^- X$, a BDT is employed with the following input variables: the invariant masses of the Higgs and $Z$ boson candidates; the angle between the two $Z$ bosons in the laboratory frame; the angles between the muons boosted in the $Z$ boson frame and the flight direction of the $Z$ boson in the laboratory frame; the angles between the $Z$ bosons boosted in the Higgs boson frame and the flight direction of the Higgs boson in the laboratory frame.
A threshold on the BDT output is determined in order to maximize the signal significance of Eq.~\ref{eq:significance}. The number of expected events after the pre-selection and the BDT threshold requirement is reported in Tab.~\ref{tab:hzz}.

As with the $H \to WW^{*}$ analysis, the counting experiment procedure is followed, assuming that the uncertainties on the efficiency and the integrated luminosity are negligible. This yields:
\begin{equation}
    \frac{\Delta \sigma (H\to ZZ^*)}{\sigma (H\to ZZ^*)} = \frac{\sqrt{S+B}}{S}\ .
\end{equation}
Finally, the estimated statistical sensitivity is:
\begin{equation}
    \frac{\Delta \sigma (H\to ZZ^*)}{\sigma (H\to ZZ^*)} = 17\%\ .
\end{equation}
%
%
\subsection[$H \rightarrow \mu^+ \mu^-$ cross section]{$H \rightarrow \mu^+ \mu^-$ cross section}
\label{sec:h2mumu}

This section presents the study of the statistical sensitivity on the $H \to \mu^+ \mu^-$ production cross section~\cite{Montella}. 
The Monte Carlo samples used in the analysis were produced with the MadGraph generator.
Two production processes of the Higgs boson are generated: $\mu^+\mu^- \rightarrow H(\rightarrow \mu^+\mu^-)\nu_\mu \bar{\nu}_\mu$ and $\mu^+\mu^- \rightarrow H (\rightarrow \mu^+\mu^-) \mu^+\mu^-$ featuring two muons and two muonic neutrinos and four muons in the final state, respectively.
The main background processes with the same final states as the sought signals are the inclusive channels $\mu^+\mu^- \rightarrow \mu^+ \mu^- \nu_\mu \bar{\nu}_\mu$ and $\mu^+\mu^- \rightarrow \mu^+\mu^-\mu^+\mu^-$, from which the contributions from the signal are removed, and $\mu^+\mu^- \rightarrow t \bar{t} \rightarrow W^+W^-b \bar{b}$ with $W^\pm \rightarrow \mu^\pm \nu_\mu (\bar{\nu}_\mu)$.
In order to reduce the computational time of the event simulation and reconstruction, the BIB was not overlaid to the physical processes, since it was shown to have a minimal impact on the muon reconstruction~\cite{epjc}. A correction was then applied to account for the BIB effects.

\begin{figure}[b]
    \centering
    \includegraphics[width=0.47\textwidth]{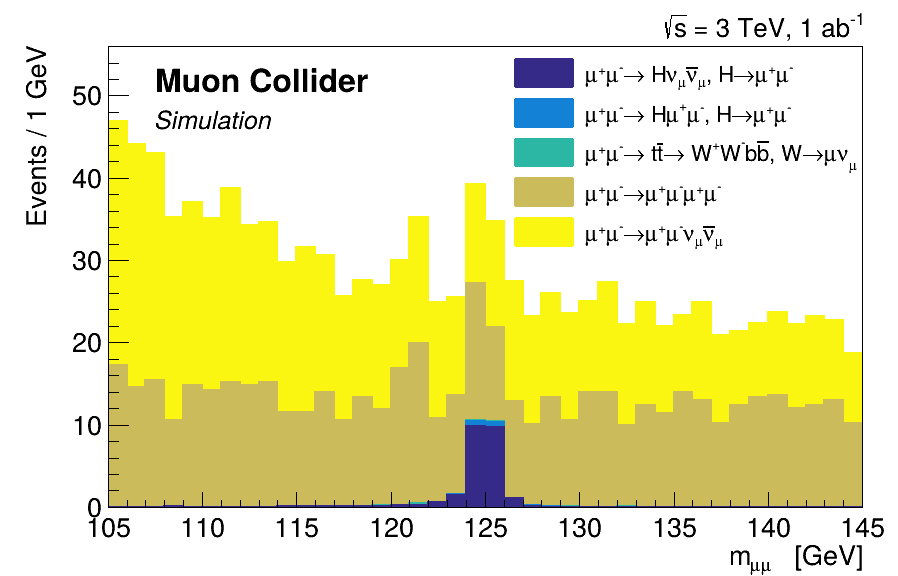}
    \caption{Stack of the $m_{\mu\mu}$ distributions for the selected signal and background events \cite{annual}. The distributions have been normalized to the expected number of events with $\mu^+\mu^-$ collisions at 3 TeV and $\mathcal{L}=1$ ab$^{-1}$, reported in Tab.\ref{tab2_hmumu}. 
    The BIB has a negligible impact on muon reconstruction and was therefore not overlaid on the physical processes.
    \label{fig2_hmumu}}
\end{figure}
\begin{table*}[t]
    \centering
    \caption{Theoretical cross section within the kinematical fiducial region, pre-selection efficiency, BDT efficiency, and expected events for signal and background processes in the $H \to \mu^+\mu^-$ analysis with $\mu^+\mu^-$ collisions at 3 TeV and $\mathcal{L}=1$ ab$^{-1}$.}
    \begin{tabular}{c|c c c c}
    \hline
    \hline
         \textbf{Process} & $\sigma \ [\mathrm{fb}]$ & $\epsilon_{presel} [\%]$ &  $\epsilon_{BDT} [\%]$& $N_{exp}$ \\
    \hline
      $\mu^+\mu^- \rightarrow H (\rightarrow \mu^+\mu^-) \nu_{\mu} \bar{\nu}_{\mu}$ & $0.109$  & 70.0 & 31.7 & 24.2 \\
      $\mu^+\mu^- \rightarrow H (\rightarrow \mu^+\mu^-) \mu^+\mu^-$                & $0.010$  & 65.2 & 25.0 &  1.6 \\
      \hline
      $\mu^+\mu^- \rightarrow \mu^+\mu^-\nu\bar{\nu}_{\mu}$                         & $11.09$  & 45.2 & 12.7 &  637 \\
      $\mu^+\mu^- \rightarrow \mu^+\mu^-\mu^+\mu^-$                                 & $297.4$ & 11.7 & 1.37 &  477 \\
      $\mu^+\mu^- \rightarrow t \bar{t} \rightarrow W^+W^- b \bar{b},$ $W \rightarrow \mu\nu_{\mu}$ & $0.320$ & 3.38 & 10.2 & 1.1 \\
      \hline
      \hline
    \end{tabular}
    \label{tab2_hmumu}
\end{table*}
The $H \to \mu^+ \mu^-$ analysis proceeds in two steps. First, a loose preselection is applied to identify events with a potential Higgs boson candidate and to suppress part of the low-energy background, then a final event selection is performed using two BDTs.
To remove a fraction of the physics backgrounds, two opposite-charge muons are required with $p_{\mathrm{T}} > 5$ GeV in the angular region $|\eta|<2.44$.   
Further requirements are applied to the dimuon invariant mass and momenta: $105 < m_{\mu\mu} < 145$ GeV, $p_{\mathrm{T}}(\mu^+\mu^-)>30$ GeV, and $p_{\mathrm{T}}(\mu^+)+p_{\mathrm{T}}(\mu^-)>50$ GeV.
%

%

After this pre-selection, a BDT classifier is trained to discriminate the total signal from the background $\mu^+\mu^- \rightarrow \mu^+ \mu^- \nu_\mu \bar{\nu}_\mu$. A second BDT classifier is trained independently for separating the signal from the background $\mu^+\mu^- \rightarrow \mu^+\mu^-\mu^+\mu^-$. The BTDs exploit observables of the two final-state muons (the $p_{\mathrm{T}}$ scalar sum, the angular separation $\Delta R = \sqrt{\Delta\phi^2 + \Delta\theta^2}$, the cosine of the angle between the three-momenta, the cosine of the angle of $\mu^-$ momentum in the dimuon rest frame with respect to the dimuon-system direction in the laboratory frame), of the dimuon system (the boost $\beta_{\mu\mu}$, the transverse momentum, the polar angle and the modulus of the three-momentum, the recoil mass),
and event observables (the total visible energy, the total visible energy minus the dimuon-system energy, the missing transverse momentum).
A cut on the BDT scores is used to select the final samples. The cut values are chosen to maximize the signal significance $\mathcal{S}$, as defined in Eq.~\ref{eq:significance}, and events are required to satisfy at least one of the two BDT requirements.
Figure~\ref{fig2_hmumu} shows the distribution of the dimuon mass $m_{\mu\mu}$ for the expected signal and background events in 1 ab$^{-1}$ of data. The corresponding event yields are listed in Tab.~\ref{tab2_hmumu}.

The number of signal events $N_S$, which is used to calculate the production cross section in Eq.~\ref{eq:nexp}, can be determined with a fit to the $m_{\mu\mu}$ distribution. The expected uncertainty on $N_S$ is estimated with a toy Monte Carlo study. The signal and background models are build from a parameterization of the distributions of Fig.~\ref{fig2_hmumu}. $N_S$ is then determined with an unbinned extended maximum-likelihood fit in 10000 pseudo experiments: the $N_S$ distribution is centered around 25.8 with an RMS of 9.9.
\begin{table*}[h!]
    \centering
    \caption{Theoretical cross section within the kinematical fiducial region, pre-selection efficiency, BDT efficiency, and expected events for signal and physics background processes that have a final state with two high energy photons in $\mu^+\mu^-$ collisions at 3 TeV and $\mathcal{L}=1$ ab$^{-1}$. $X$ indicates both $\nu\bar{\nu}$ and $\ell\ell$.}
    \begin{tabular}{c|c c c c}
    \hline
    \hline
         \textbf{Process} &  $\sigma \ [\mathrm{fb}]$& $\epsilon_{presel} [\%]$ & $\epsilon_{BDT} [\%]$& $N_{exp}$ \\
    \hline
       $\mu^+ \mu^- \to H (\to \gamma \gamma) X $ & $0.91$ & $86.1$ & $50.6$ & $396$ \\
       \hline
        $\mu^+ \mu^- \to \nu\bar{\nu} \gamma \gamma$ & $81.98$ & $53.9$ & $1.0$ & $442$ \\
        $\mu^+ \mu^- \to \ell\ell \gamma \gamma$ & $4.41$ & $76.4$ & $0.32$ & $11$ \\
        $\mu^+ \mu^- \to \ell\ell \gamma$ & $159.01$ & $33.0$ & $0.06$ & $31$ \\
        $\mu^+ \mu^- \to \gamma \gamma$ & $60.15$ & $99.1$ & $0.02$ & $12$ \\
      \hline
      \hline
    \end{tabular}
    \label{tab:h2gammagamma}
\end{table*}
To assess the impact of the BIB, the signal sample $\mu^+\mu^- \rightarrow H \nu_{\mu} \bar{\nu}_{\mu}$ was reconstructed with the BIB superimposed and compared to the case without BIB at the preselection level of the analysis. The $H \rightarrow \mu^+ \mu^-$ selection efficiency results reduced by about 2\%, while the width of the dimuon mass peak degrades only by a few percent, which is within the reached precision. 
Assuming negligible uncertainties on the selection efficiency and on the integrated luminosity, the relative uncertainty on $N_S$, corrected for the BIB effects, can be considered as the statistical uncertainty of the cross-section measurement:
\begin{equation}
    \frac{\Delta \sigma (H \rightarrow \mu^+ \mu^-)}{\sigma (H \rightarrow \mu^+ \mu^-)} = 39\% \ .
\end{equation}
It is worth noting that the possibility of identifying the $ZZ$-fusion processes by detecting the forward-scattered muons,
would significantly reduce the $\mu^+\mu^- \rightarrow \mu^+\mu^-\mu^+\mu^-$ background and improve the cross-section sensitivity.
%
%
\subsection[$H \to \gamma \gamma$ cross section]{$H \to \gamma \gamma$ cross section}
\label{sec:h2gammagamma}


\begin{figure}[b]
\centering
    \includegraphics[width=0.95\linewidth]{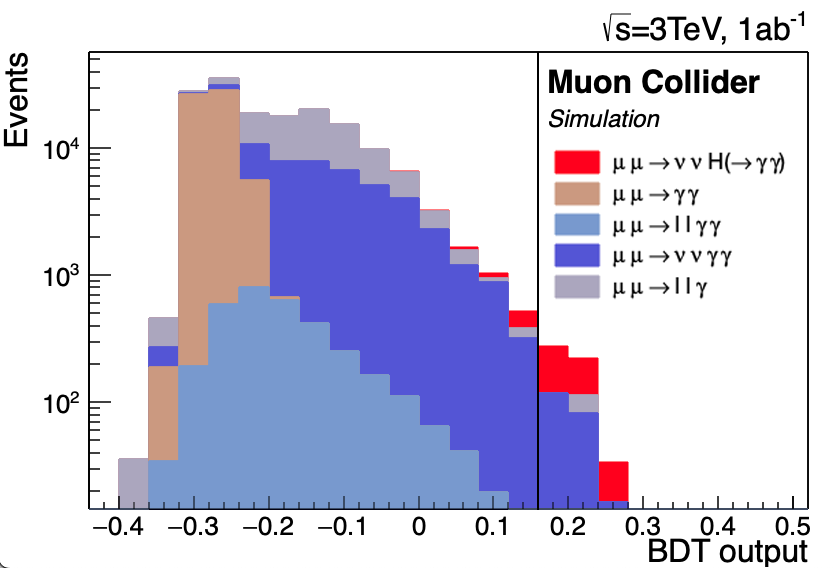}
    \caption{
    Distribution of the BDT output of the signal and major physics background  process. The distributions are stack and normalized to the expected number of events with $\mu^+\mu^-$ collisions at 3 TeV and $\mathcal{L}=1$ ab$^{-1}$ after the preselection cuts. The black vertical line indicates the value of the BDT cut.}
    \label{fig:bdt_hgammagamma}
\end{figure}
%
The experimental signature for the process $\mu\mu\to H(\to\gamma\gamma)X$ consists of two high-$p_{\mathrm{T}}$ photons with an invariant mass compatible with the Higgs boson mass. 
The signal and physics background processes are generated with MadGraph. The list of the considered backgrounds and their production cross sections is reported in Tab.~\ref{tab:h2gammagamma}, where only processes with photons in the final state and the highest cross sections are taken into account. The processes with leptons in the final state can pass the event selection because of photon misidentification. 

Events are selected with at least two reconstructed photons having $E>15$ GeV, $p_{\mathrm{T}}>10$ GeV, and $|\eta|<2.436$. The most energetic photon must have a $p_{\mathrm{T}}>40$ GeV. If more than two photons in the event fulfill these requirements, the two with the highest energy are selected. 
The invariant mass of the two photons must satisfy the condition $m_{\gamma\gamma}>40$ GeV. 


\begin{figure}[b]
  \centering
        \includegraphics[width=0.95\linewidth]{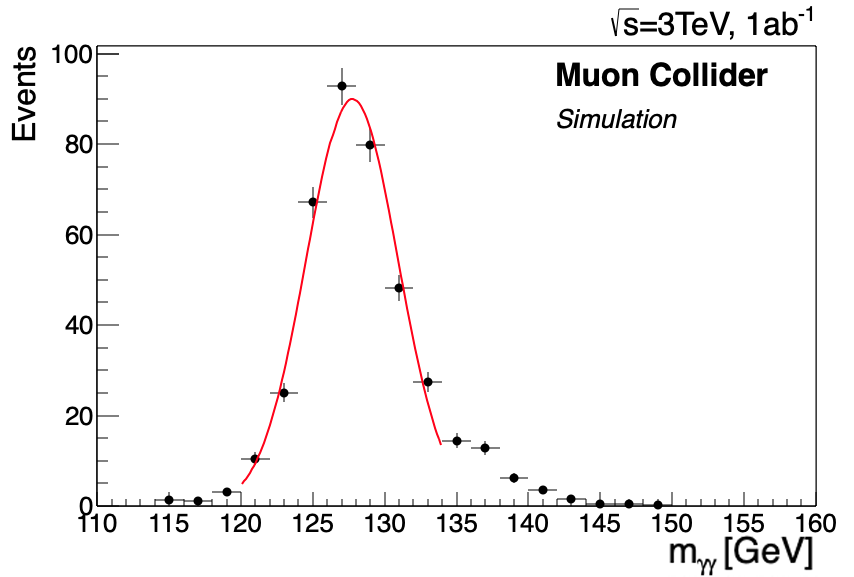}
        \caption{Distribution of reconstructed $m_{\gamma\gamma}$ for the selected $H \to \gamma \gamma$ events, with no physics background, overlaid with a Gaussian fit. The distribution is normalized to the expected number of events with $\mu^+\mu^-$ collisions at 3 TeV and $\mathcal{L}=1$ ab$^{-1}$.}
        \label{fig:mass_fit}
\end{figure}
After the preselection, signal and background events are classified using a BDT. The following variables are used for the classification: the invariant mass $m_{\gamma\gamma}$ and the transverse momentum $p_{\mathrm{T}}(\gamma\gamma)$ of the di-photon system; $p_{\mathrm{T}}$ of the two photons; $\Delta \theta$ and $\Delta \phi$ between the two photons; the acollinearity between the two photons (Eq.~\ref{eq:acol}); the acollinearity between each photon and the Higgs candidate.
The BDT is trained by using the signal sample and a mixture of the background samples, merged together with equal weights.
The distributions of the BDT output are shown in Fig.~\ref{fig:bdt_hgammagamma}. 
A cut  on the BDT output is applied to maximize the significance of Eq.~\ref{eq:significance}.
The expected number of signal and background events after the BDT requirement are reported in Tab.~\ref{tab:h2gammagamma}. 
Fig.~\ref{fig:mass_fit} shows the invariant mass of the Higgs candidates, with no physics background, after the preselection and the BDT cut. A Gaussian fit of the invariant mass distribution gives a mass resolution of 3.2 GeV for the signal sample.

\begin{table*}[h!]
    \centering
    \caption{Comparison of statistical sensitivities for Higgs boson production cross sections at a $\sqrt{s}=3$ TeV Muon Collider with $\mathcal{L} = 1$ ab$^{-1}$, obtained with a detailed detector simulation with the BIB overlaid and without the BIB (target performance).}
    \begin{tabular}{c|c|c}
    \hline
    \hline
         \textbf{Channel} &  Detailed simulation with BIB& Target performance \\
          &  $\Delta \sigma/\sigma$ [\%] & $\Delta \sigma/\sigma$ [\%] \\
    \hline
       $H \to b \bar{b}$ &  0.78 & 0.55 \\
      $H \to WW^*$ & 2.9 & 2.0 \\
      $H \to ZZ^*$ & 17 & 12 \\
      $H \to \mu^+\mu^-$ & 39 & 38 \\
      $H \to \gamma \gamma$ & 7.5 & 7.0 \\
      \hline
      \hline
    \end{tabular}
    \label{tab:comparison_tab}
\end{table*}
Assuming that the signal yield is obtained by subtracting the expected number of background events from the observed number of events, after the BDT cut, and assuming a negligible uncertainty on the luminosity and on the signal selection efficiency, the statistical sensitivity on the measurement of the $H \to \gamma \gamma$ cross section can be evaluated as:
\begin{equation}
    \frac{\Delta \sigma (H\to \gamma \gamma)}{\sigma (H\to \gamma \gamma)} = \frac{\sqrt{S+B}}{S}\ .
\end{equation}
With the expected number of events reported in Tab.~\ref{tab:h2gammagamma}, the estimated statistical sensitivity is:
\begin{equation}
    \frac{\Delta \sigma (H\to \gamma \gamma)}{\sigma (H\to \gamma \gamma)} = 7.5\%\ .
\end{equation}

\subsection{Comparison of results from detailed simulation and assumed target detector performance}
\label{sec:parametric}

The results presented in the preceding sections are compared with those obtained under the ideal condition of no BIB in the detector. While the BIB will always affect the detector, future advancements in technologies for tracker devices and calorimeter systems are expected to significantly reduce its impact. These potential improvements can now be accounted for by assuming that the effects of BIB are negligible. This can be considered a conservative assumption given the efforts involved in detector R\&D.

For each of the Higgs boson decay channels presented in the previous section, signal and physics background events are reconstructed without overlaying the BIB. The pre-selection requirements were redefined, and the BDTs were re-trained accordingly. This is necessary because the BIB represents the primary constraint for event reconstruction algorithms and analysis strategy optimization.
The comparison is presented in Tab.~\ref{tab:comparison_tab}.


It should be noted that in the cases of $H \to \mu^+\mu^-$ and $H \to \gamma \gamma$ analyses the cross section uncertainties are similar. This is expected, as the BIB contribution to the reconstruction of high-energy photons and muons is negligible.
For $H \to b \bar{b}$, $H \to WW^*$ and $H \to ZZ^*$, the cross section uncertainties with the detailed simulation are about 30\% higher than the target ones: the reason is that the jet reconstruction algorithms are not yet fully optimized for handling the BIB. Ongoing developments are expected to significantly reduce this gap.

\section{Higgs boson width}
\label{sec:width}

The Higgs width ($\Gamma_H$) is a fundamental parameter in Higgs boson physics, since it is closely linked to the broader understanding of Higgs couplings. The expected value in the SM is $\Gamma_H= 4.07$ MeV for a 125 GeV Higgs boson~\cite{ref:pdg}, a deviation from this value would be a clear indication of new physics in the Higgs sector. Moreover, a model-independent measurement of $\Gamma_H$ is an essential input to the Higgs couplings fit, which treats the probability of Higgs boson decays to invisible final states as a free parameter.

At $e^+e^-$ colliders at low center-of-mass energy ($\sqrt{s}=240$ GeV), the $e^+e^- \to HZ$ process with full reconstructed $Z$ bosons allows for the determination of the Higgs boson kinematics independently of the Higgs decay channel~\cite{higgs_report}. The advantage of this approach is that it is based on minimal assumptions.

\begin{figure}[b]
    \centering
    \includegraphics[width=0.45\textwidth]{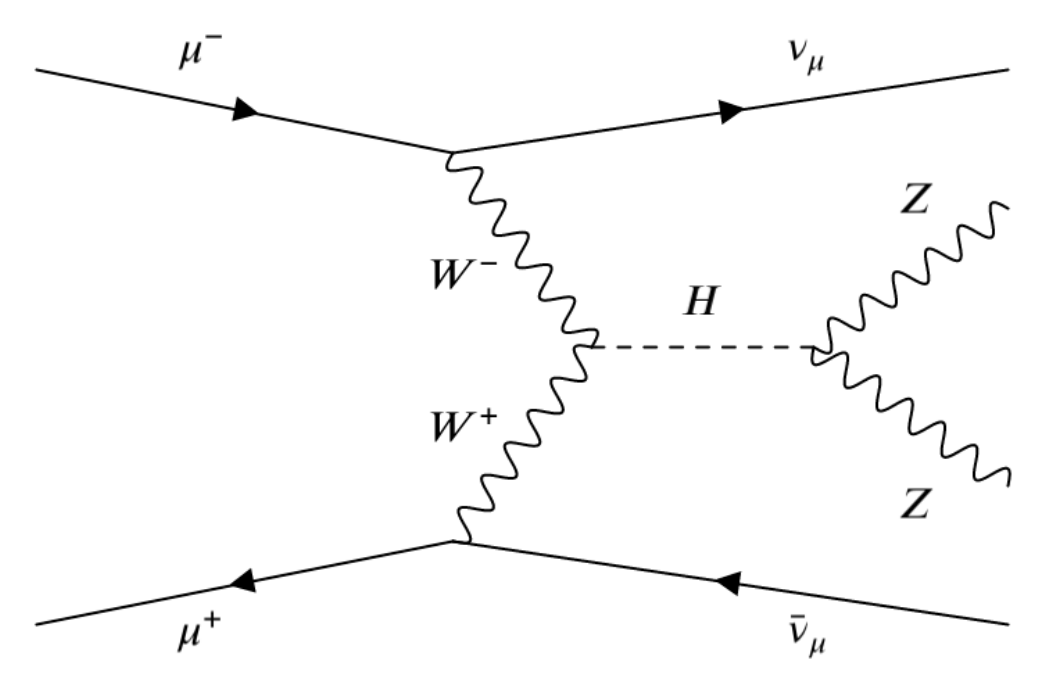}
    \caption{Feynman diagram for on-shell $ZZ$ production via VBF, occurring with an off-shell Higgs propagator.}
    \label{fig:Hoffs-diagram}
\end{figure}
At a 3~TeV MuC, the $HZ$ production cross section is relatively small (see Fig.\ref{fig:HXS}), therefore the proposed method for the $e^+e^-$ colliders is not a viable option. Instead, the approach considered for HL-LHC, where $\Gamma_H$ is obtained by studying the kinematics of the full $ZZ$ process \cite{higgs_report}, is used in this paper~\cite{GiambastianiPhD}. This is a model-dependent approach, based on the interference between the off-shell Higgs boson production and the full amplitude in $ZZ\to 4 \ell$. Hence, the pair production of on-shell $Z$ bosons via the VBF process $\mu^+ \mu^- \rightarrow Z Z \nu_{\mu} \bar\nu_{\mu}$ is simulated in the detector and reconstructed for the determination of $\Gamma_H$.


This final state has a contribution coming from the decay of an off-shell Higgs boson, as shown in Fig.~\ref{fig:Hoffs-diagram}, therefore the process is sensitive to the couplings of the Higgs boson to $Z$ bosons.
Thanks to the contribution of this process, the production of a pair of $Z$ bosons can be used directly to extract the modifier $\kappa$ of the couplings between Higgs boson and vector bosons, assuming it to be the same for both $Z$ and $W^{\pm}$:
\begin{equation}
	\kappa = \frac{g_{HZZ}}{g_{HZZ}^{SM}} = \frac{g_{HWW}}{g_{HWW}^{SM}}\ .
\end{equation}
To obtain $\Gamma_H$, the measurement of the $H \rightarrow ZZ^*$ cross section is also needed (Sec.~\ref{sec:h2zz}). Indeed, the following equation holds \cite{Li:2024joa}:
\begin{equation}
	\label{eq:sigmah2zz}
	\frac{\sigma(H \rightarrow Z Z^*)}{\sigma^{SM}(H \rightarrow Z Z^*)} \propto \frac{\kappa^4}{\Gamma_H/\Gamma_H^{SM}}\ ,
\end{equation}
where $\sigma^{SM}(H \rightarrow Z Z^*)$ and $\Gamma_H^{SM}$ are the expected SM values for the $W^+ W^- \rightarrow H \rightarrow Z Z^*$ cross section and Higgs boson width, respectively.
With $\kappa$ obtained from the $Z Z$ production and the measurement of the $ H \rightarrow Z Z^*$ cross section (Sec.~\ref{sec:h2zz}), the determination of $\Gamma_H$ is finally possible.
%
%
The approach for measuring $\kappa$ consists in comparing data samples generated by varying $\kappa$, and then determining the best $\kappa$ value through a likelihood scan. 

The $\mu^+ \mu^- q\bar{q}$ final state, where $q$ is any quark, has been chosen as a compromise between event counts and precision in the reconstruction of the final state.
The relevant processes come from VBF di-boson production with the $\mu^+ \mu^- q\bar{q}$ final state, namely: $\mu^+ \mu^- \rightarrow \nu_{\mu} \bar\nu_{\mu} Z(\rightarrow \mu^+ \mu^-) Z(\rightarrow q\bar{q})$, $\mu^+ \mu^- \rightarrow \nu_{\mu} \bar\nu_{\mu} H(\rightarrow b \bar b) Z(\rightarrow \mu^+ \mu^-)$, and $\mu^+ \mu^- \rightarrow \nu_{\mu} \mu^{\pm} W^{\pm}(\rightarrow q\bar{q}) Z(\rightarrow \mu^+ \mu^-)$.
The first two are $WW$-fusion processes and are the ones with the largest dependence on $\kappa$. The latter is a $WZ$-boson fusion process, where the first muon is usually undetected, since it is emitted at very small polar angles. In Ref.~\cite{Forslund:2023reu}, it has been verified that the $ZZ$-fusion contribution to the $\Gamma_H$ sensitivity is negligible, therefore it is not considered in this analysis.

Several high-statistics simulation samples corresponding to different $\kappa$ values should be produced for determining the uncertainty on $\kappa$. The task of performing the detailed simulation for all these samples is excessively computing intensive and time demanding with the resources available. For this reason a hybrid fast-simulation/full-simulation approach has been used: 
\begin{itemize}
\item A sample of $\mu^+ \mu^- q\bar{q}$ events has been generated with Madgraph with $\kappa=1.0$ and processed with the detailed detector simulation.
This sample has been used to determine the reconstruction efficiencies and momentum resolutions for jets and muons, as well as their dependence on the kinematics.
\item $\mu^+ \mu^- q\bar{q}$ samples have been generated with Madgraph, corresponding to 17 values of $\kappa$ ranging form $-0.2$ to $2.0$, to which the resolutions and efficiencies obtained with the detailed simulation have been applied.
Resolutions are applied as smearing factors and efficiencies as event weights. The dependence on the kinematics is taken into account.
\item These fast-simulation samples are then used to perform the likelihood scan for the determination of the uncertainty on $\kappa$.
\end{itemize}

\begin{figure}[t]
	\centering
	\includegraphics[width=0.48\textwidth]{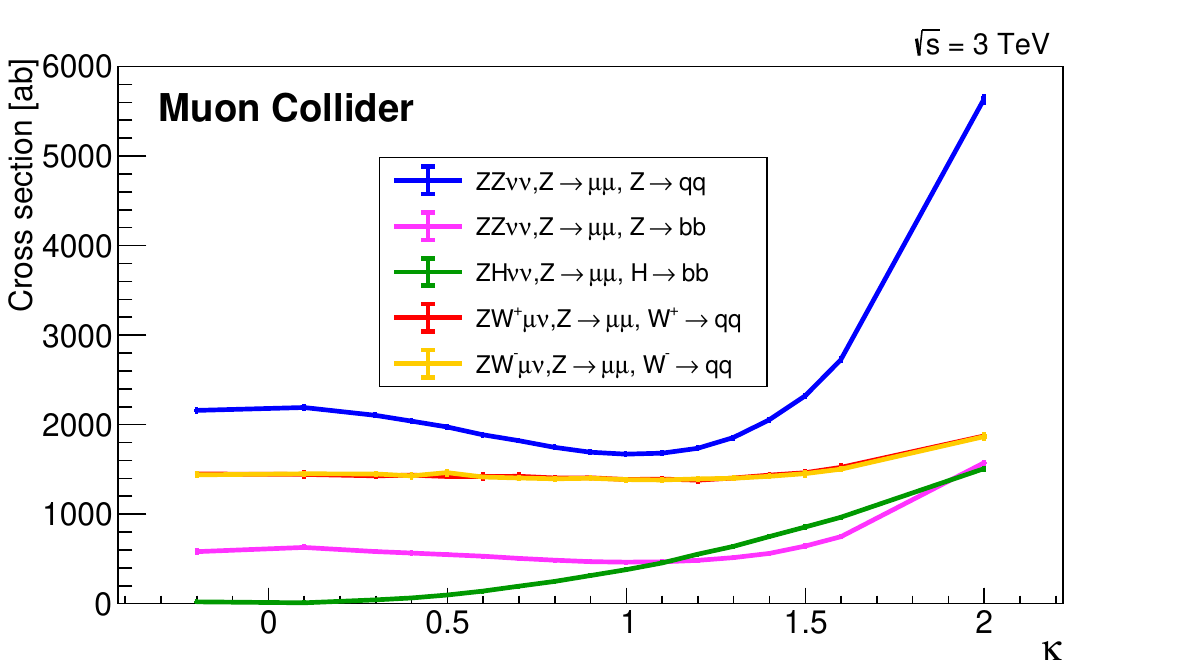}
	\caption{Cross sections obtained with Madgraph for the processes considered in the determination of the Higgs coupling modifier $\kappa$ for $Z$ and $W$ bosons.}
	\label{fig:xsecs}
\end{figure}

The cross sections, calculated with Madgraph, of the main processes that contribute to $\mu^+\mu^- q\bar{q}$ are shown in Fig.~\ref{fig:xsecs} as a function of $\kappa$.
It is evident that $\mu^+ \mu^- \rightarrow \nu_{\mu} \bar\nu_{\mu} Z(\rightarrow \mu^+ \mu^-) Z(\rightarrow q\bar{q})$ is the most sensitive process to $\kappa$.
In this analysis, jets and muons are reconstructed as in Sec.~\ref{sec:h2zz}.
Events are selected requiring at least two opposite-charge muons with $p_{\mathrm{T}}>15$ GeV and at least two jets with $p_{\mathrm{T}}>20$ GeV. The positive and negative muons with the highest $p_{\mathrm{T}}$ are selected and required to have an invariant mass between 85 and 97 GeV. Jets with a $\Delta R < 0.5$ with respect to any of the two selected muons are discarded. 
To remove fake jets from the BIB, jets are required to contain at least one track. After this requirement fake jets are assumed to be negligible and they are not considered in the analysis. 
Finally, the two jets with the highest $p_{\mathrm{T}}$ are selected and required to have an invariant mass greater than 50 GeV. The invariant mass $M_{\mu\mu qq}$ of all selected final state objects ($\mu^+$, $\mu^-$ and the two jets) is required to be between 180 and 1800 GeV.

\begin{figure}[t]
	\centering
	\includegraphics[width=0.48\textwidth]{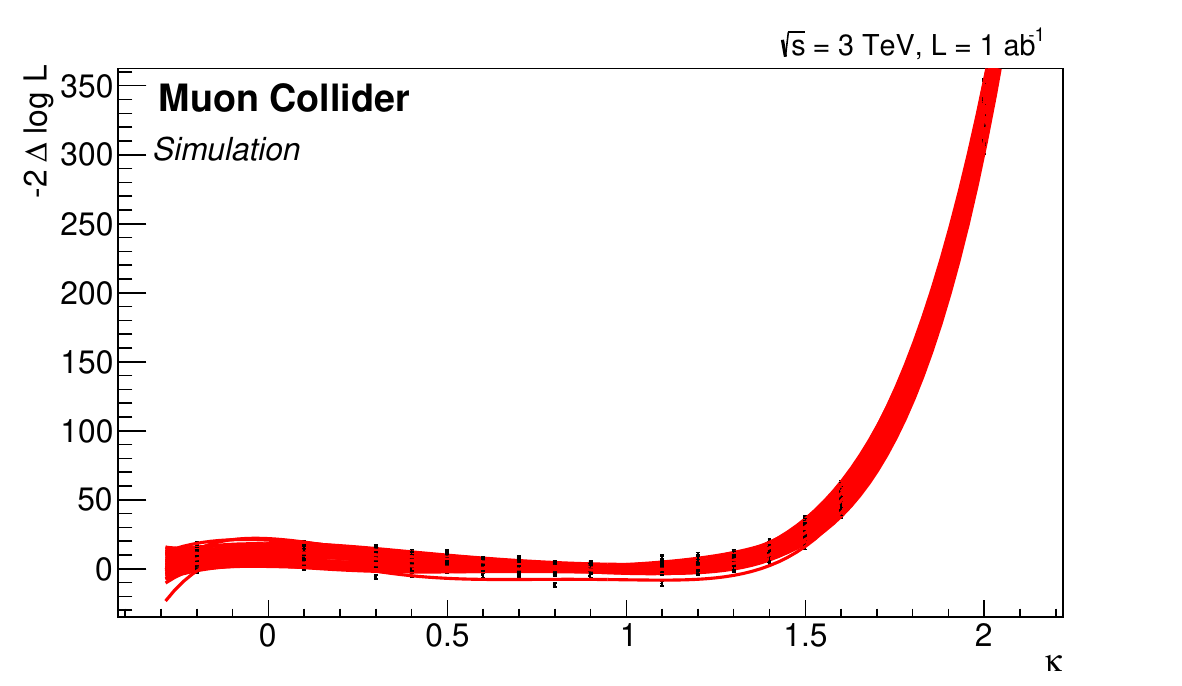}
	\caption{Likelihood for the determination of $\kappa$, fitted with a 6$^{\mathrm{th}}$ degree polynomial. Likelihood scans for 50 pseudo-experiments are overlaid (red lines), the points represent the average and standard deviations of $-2\Delta \log(L)$ for the considered $\kappa$ values. This result has been obtained with a hybrid fast simulation/detailed simulation described in the text, that takes into account the BIB impact.}
	\label{fig:scans}
\end{figure}

Likelihood values are calculated comparing 2-dimensional distributions of $M_{\mu\mu qq}$ and $\theta_{ZZ}$ (\emph{i.e.} the angle between the reconstructed $Z$ bosons in the laboratory frame) between pseudo-dataset and simulation distributions. Pseudo-datasets are generated from the distribution with $\kappa = 1$. For each pseudo-dataset, the likelihood difference between $\kappa=1$ and a $\kappa$-hypothesis, $-2\Delta \log(L)$ as a function of $\kappa$ is obtained. These values are fitted to a 6$^{\mathrm{th}}$ degree polynomial, and the $1\,\sigma$ interval is estimated as the interval around $\kappa=1$ where the fitted polynomial has a value below 1. 
For each pseudo-dataset, upper and lower bounds are found, and the final result is given by averaging the intervals obtained with 50 different generated pseudo-datasets.
The obtained likelihood scans are shown in Fig.~\ref{fig:scans}.
The final result on the $\kappa$ uncertainty interval is $\kappa = 1^{+17\%}_{-12\%}$.

\begin{table*}[h!]
    \centering
    \caption{Theoretical cross section within the kinematical fiducial volume, pre-selection efficiency, $b$-tagging efficiency, and expected events for signal and background processes in the $HH \rightarrow b\bar{b} b\bar{b}$ analysis with $\mu^+\mu^-$ collisions at 3 TeV and $\mathcal{L}=1$ ab$^{-1}$. $X$ indicates both $\nu\bar{\nu}$ and $\ell\ell$.}
    \begin{tabular}{c|c c c c}
    \hline
    \hline
         \textbf{Process} & $\sigma \ [\mathrm{fb}]$ &  $\epsilon_{presel} [\%]$ &  $\epsilon_{tag} [\%]$ & $N_{exp}$ \\
    \hline
       $\mu^+ \mu^- \to H H X \to b \bar{b} b \bar{b} X$         & 0.31 & $38.8$ &$69.8$ & $84$ \\
       \hline
       $\mu^+ \mu^- \to H (\to b \bar{b}) q_{h}  \bar{q}_{h} (\nu\nu, \ell \ell)$  & 3.1  & $39.2$ &$ 62.6$ & $761$ \\
       $\mu^+ \mu^- \to q_{h}  \bar{q}_{h} q_{h}  \bar{q}_{h} (\nu\nu, \ell \ell)$ & 5.9  & $36.5$ &$49.5$ & $1066$ \\
       
      \hline
      \hline
    \end{tabular}
    \label{tab:HH_negl_mistag}
\end{table*}
%


Starting from Eq.~\ref{eq:sigmah2zz}, it is possible to derive:
\begin{equation}
	\Delta \Gamma_H = \Delta\sigma(H \rightarrow Z Z^*) \oplus (4 \, \Delta \kappa)\ ,
\end{equation}
where the $\oplus$ symbol means squared sum. Taking $\Delta\sigma(H \rightarrow Z Z^*)$ from Sec.~\ref{sec:h2zz}, the final result is found to be
\begin{equation}
  \frac{\Gamma_H}{\Gamma_H^{SM}} =  1^{+71\%}_{-88\%}\ .  
\end{equation}
The result for the Higgs width is compared to the phenomenological study presented in Ref.~\cite{Forslund:2023reu}.
In the article, the precision on the Higgs width, obtained with a fully general $\kappa$-fit, is 24$\%$.
However, the contribution obtained by exploiting only the $\mu^+ \mu^- \rightarrow \nu_{\mu} \bar\nu_{\mu} ZZ$ and $\mu^+ \mu^- \rightarrow \nu_{\mu} \bar\nu_{\mu} W^+W^-$ channels is 58$\%$~\cite{Matttalk}. The latter result includes different final states ($4q$, $2q2l$, and $l \nu qq$) and is of the same order as the one presented in this section ($\Delta \Gamma_H \sim 80 \%$), which uses only $2q2 \mu$ events and is based on the detailed detector simulation. 
It is worth mentioning that the aforementioned article quotes an uncertainty of $\Delta \Gamma_H = 3.4 \%$ for muon collisions at $\sqrt{s}=10$ TeV. The limitation of MuC at at $\sqrt{s}=3$ TeV could be overcome by using higher energy muon collisions.

\section{Trilinear coupling}
\label{sec:double}

The state-of-the-art projections on the $\kappa_{\lambda_3}$ sensitivity obtained with MuC parametric simulations are reported in Refs.~\cite{3tevphysics,epjc}. In those studies, the 68$\%$ confidence-level bound, obtained with $\mathcal{L} = 1$ ab$^{-1}$ of data collected at $\sqrt{s}=3$ TeV, is composed of two intervals, corresponding to two different likelihood minima: $0.73 < \kappa_{\lambda_3} < 1.35$ and $1.85 < \kappa_{\lambda_3} < 1.94$. It is also stated that the second minimum can be removed by collecting $\mathcal{L} = 2$ ab$^{-1}$, leading to a 68$\%$ confidence level of $0.85 < \kappa_{\lambda_3} < 1.16$. In the following, the case with $\mathcal{L} = 1$ ab$^{-1}$ is considered for the detailed simulation, focusing on the first minimum, which is closer to the SM value.

Sec.~\ref{sec:HH} describes the measurement of the double Higgs production ($HH$), which is the first step towards a measurement of the trilinear coupling. Then, the trilinear coupling extraction is presented in Sec.~\ref{sec:trilinear}.

\subsection[$HH \rightarrow b\bar{b} b\bar{b}$ cross section]{$HH \rightarrow b\bar{b} b\bar{b}$ cross section determination}
\label{sec:HH}

The $HH$ production is studied using the $H \to b \bar{b}$ decay for both Higgs bosons, since this final state exhibits the highest branching ratio, and the channel with four $b$ quarks provides the highest statistics~\cite{Buonincontri,BuonincontriPhD}.
The contribution of other final states, like $HH \to b \bar{b}W^+W^-$, will be investigated in future studies.


In this analysis, at least four reconstructed jets are required in the event with a minimum $p_{\mathrm{T}}$ of 20 GeV. Input tracks for jets are obtained with the CKF algorithm.
The $b$-jet identification efficiencies and the misidentification rates presented in Sec.~\ref{sec:eventreco}, have been applied to the reconstructed jets as a function of the jet $p_{\mathrm{T}}$ and taking into account the flavour composition of the final state, as with the $H \to b \bar{b}$ analysis in Sec.~\ref{sec:h2bb}. 

To reconstruct the $HH$ events, all possible two-jet combinations are formed, in which at least one jet is requested to be identified as a $b$-jet.
Two Higgs boson candidates are then built from the two jet pairs whose invariant masses $m_{12}$ and $m_{34}$ minimize the figure of merit:
\begin{equation}
F=\sqrt{(m_{12}-m_{H})^2+(m_{34}-m_{H})^2}\ ,
\end{equation}\label{F_figure_merit}
where $m_H$ is the nominal Higgs boson mass. 

The main physics background contribution comes from processes with four heavy-quark jets in the final state, $\mu^+ \mu^- \to q_h \bar{q}_h q_h \bar{q}_h (\nu\nu, \ell \ell)$  ($q_h=b$ or $c$), which comprise multiple intermediate electroweak gauge bosons. 
The other important source of background is the process $\mu^+ \mu^- \to H  q_h \bar{q}_h (\nu\nu, \ell \ell)\to b \bar{b}  q_h \bar{q}_h (\nu\nu, \ell \ell)$ where the $HHH$ vertex is not present. 
Physics backgrounds resulting from light quarks and fake jets are assumed negligible. The rationale for this assumption is that advanced tagging methods, based on machine learning techniques, should be capable of rejecting these events while keeping the signal, as discussed later in Sec.~\ref{sec:trilinear}. 
For this reason the $\mu \mu \to q\bar{q}q\bar{q}\nu \ell$ background is considered negligible. It has been also checked that the $\mu \mu \to q\bar{q}q\bar{q}$ does not significantly enter the analysis.
A summary table presenting the considered signal and background processes, along with their cross sections and expected event yields, is provided in Tab.~\ref{tab:HH_negl_mistag}.


       

\begin{figure}[b]
  \centering
    \includegraphics[width=0.95\linewidth]{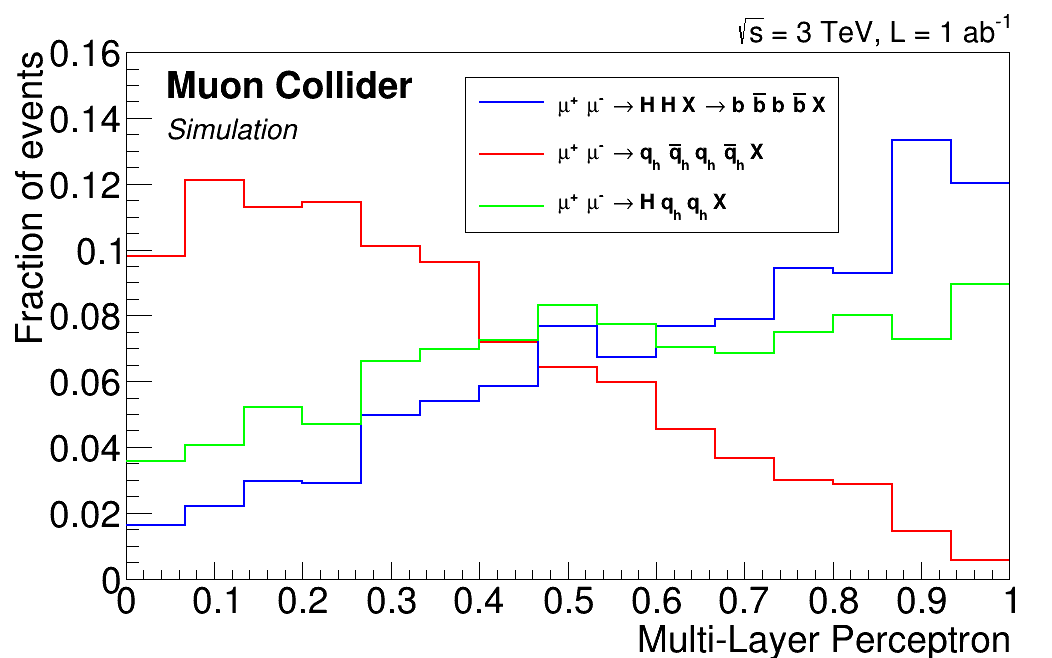}
    \caption{Distributions of the MLP output for the $HH$ signal and the main background contributions. The distributions are normalized to unit area. In the legend, $X$ stands for both $\nu\bar{\nu}$ and $\ell\ell$.}
    \label{fig:mlp}
\end{figure}
\begin{table*}[h!]
    \centering
    \caption{Theoretical cross section values, selection efficiencies, and expected number of $HH \to b\bar{b}b\bar{b}$ events for all $\kappa_{\lambda_3}$ hypotheses, assuming 1 ab$^{-1}$ of integrated luminosity at $\sqrt{s}=3$ TeV. }
    \begin{tabular}{c|c|c|c}
    \hline\hline
    $\kappa_{\lambda_3}$   &  $\sigma$ [fb] & $\epsilon$ [\%]& N$_{events}$ \\
    \hline             
          0.2   & 0.52 &30.1& 169 \\
          0.4   & 0.48 &29.3 &141  \\
          0.6   & 0.41 &28.1 &116 \\
          0.8   & 0.35 & 27.5 &97 \\ 
          0.9   & 0.33 &27.6 &92 \\   
          1.0   & 0.31 & 27.2 &84 \\  
          1.1   & 0.29 & 25.7 &74 \\            
          1.2   & 0.27 & 25.7 &71 \\
          1.4   & 0.25 & 26.5 &68 \\
          1.6   & 0.24 &25.1 &61 \\
          1.8   & 0.25 & 24.8 &61 \\
    \hline
    \hline
    \end{tabular}
    \label{tab:kappa_values}
\end{table*}
A Multilayer Perceptron (MLP)~\cite{hoecker2009tmva} has been trained on twelve observables to separate the signal events from the physics background $\mu^+ \mu^- \to q_{h}  \bar{q}_{h} q_{h}  \bar{q}_{h} X$: 
\begin{itemize}
\item the invariant masses of the jet pairs with the highest (leading candidate) and the lowest (sub-leading candidate) transverse momentum;
\item the module of the vectorial sum of the four jet momenta;
\item the sum of the four jet energies; 
\item the angle between the two jets relative to the leading candidate;
\item the maximum separation angle between the jets in the event;
\item the angle between the highest-$p_{\mathrm{T}}$ jet in the pair relative to the leading candidate, and the angle between the highest-$p_{\mathrm{T}}$ jet in the pair relative to the sub-leading candidate, with respect to the $z$-axis;
\item the four jet transverse momenta. 
\end{itemize}
The MLP output distributions for the $HH$ signal and background samples are shown in Fig.~\ref{fig:mlp}. It can be observed that a good separation between the $HH$ and $q_{h}  \bar{q}_{h} q_{h}  \bar{q}_{h} X$ processes is achieved. It has been verified that an additional MLP to separate the signal from the $\mu^+ \mu^- \to H (\to b \bar{b}) q_{h}  \bar{q}_{h} X$ background does not improve significally the measurement, since the kinematic properties of this process are very similar to those of the signal.

Using the MLP output distributions, pseudo-datasets have been generated according to the expected number of events. The yield of the $HH$ signal is extracted by fitting the MLP distribution: an average statistical uncertainty of approximately 33\% is obtained on the signal yield. The $HH$ cross section is calculated with Eq.~\ref{eq:nexp}. Therefore, assuming negligible uncertainties on the selection efficiency and integrated luminosity, the statistical uncertainty on the double-Higgs production cross section is:
\begin{equation}
    \frac{\Delta \sigma (HH \rightarrow b\bar{b} b\bar{b})}{\sigma (HH \rightarrow b\bar{b} b\bar{b})} = 33\%\ .
\end{equation}

At this time, there is no estimate of the statistical sensitivity on the $HH$ cross section based on a parametric simulation of the detector but only on the trilinear coupling with no 
detector effects~\cite{Han_2021}.


\subsection{Trilinear Higgs coupling precision determination}
\label{sec:trilinear}

\begin{figure*}[h!]
\centering
\includegraphics[width=0.46\textwidth]{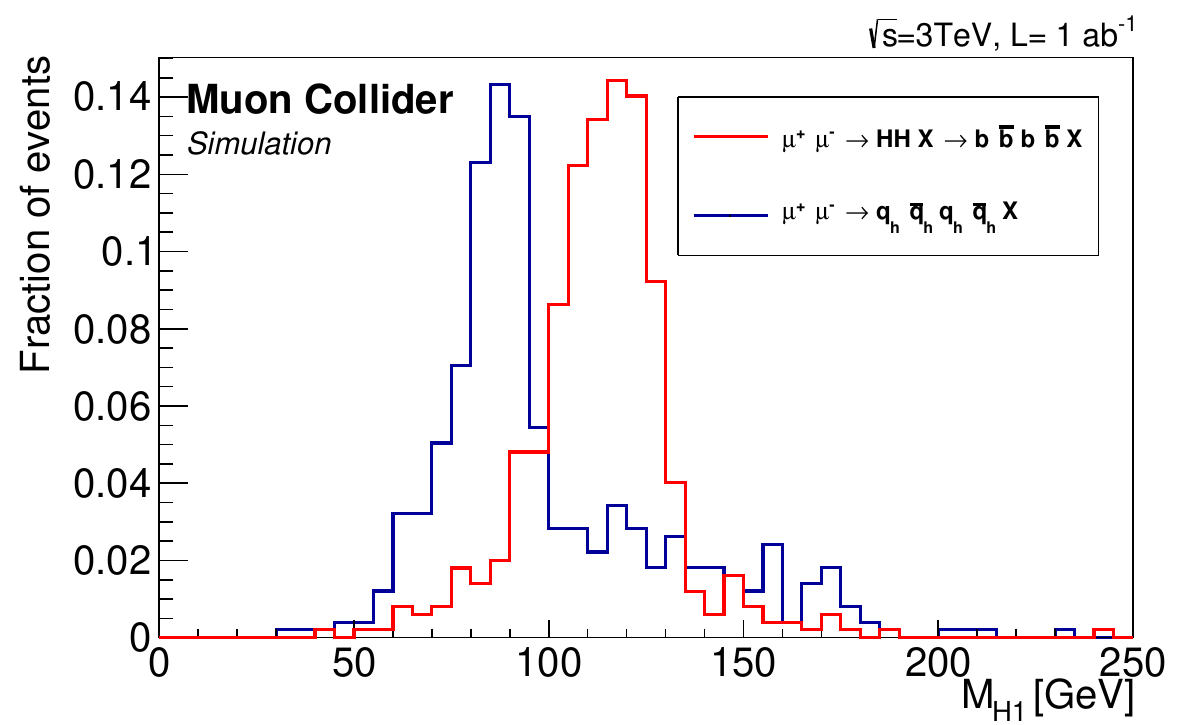}
\includegraphics[width=0.47\textwidth]{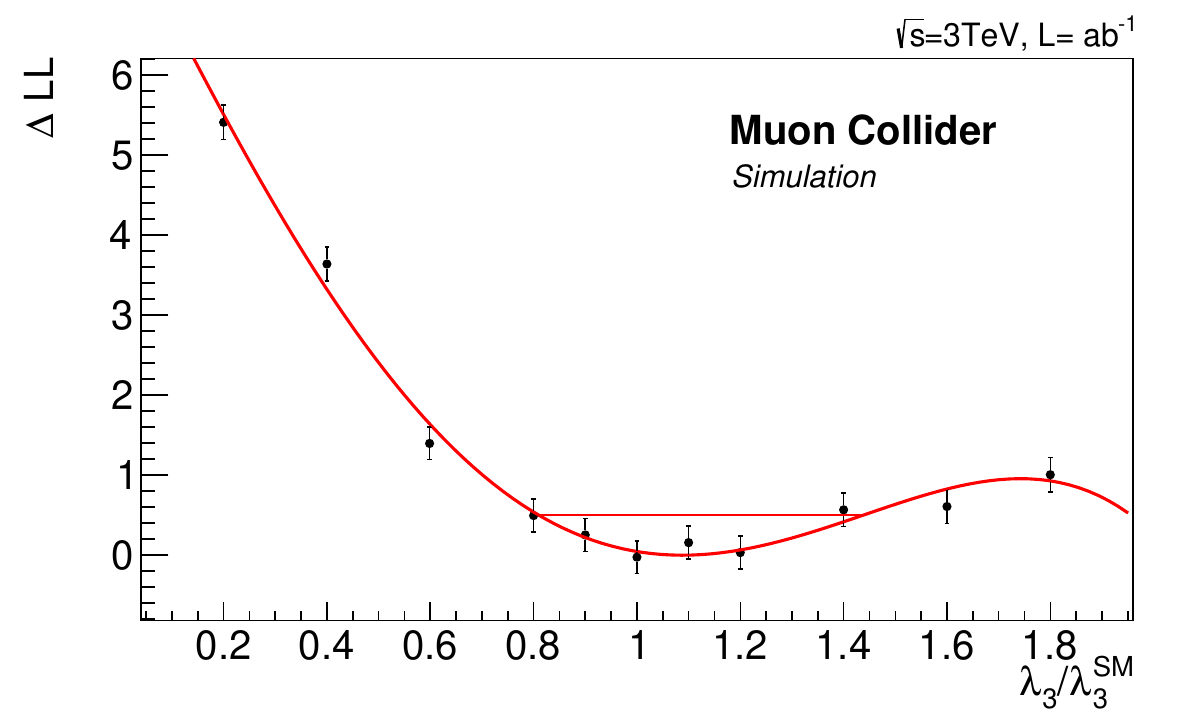}
\caption{Left: Invariant mass of the leading Higgs candidate compared between $HH$ events and $q_h \bar{q}_h q_h \bar{q}_h$ events, with samples reconstructed without the BIB. It is recalled here that the contribution from $q_h\bar{q} H$ is not included in this sample. In the legend, $X$ stands for both $\nu\bar{\nu}$ and $\ell\ell$. Right: $\Delta LL$ as a function of $\kappa$ hypothesis for samples reconstructed without the BIB, selected by assuming LHCb tagging performance ($b$-tagging efficiency of 76$\%$ and $c$-mistag rate of 20$\%$), by requiring three tagged jets in the event.} \label{inv_mass_wo_BIB}
\end{figure*}
%
The total double-Higgs production cross section is sensitive to the value of the trilinear self-coupling: Fig.~\ref{fig:HHFey} shows the three Feynman diagrams that primarily contribute to the double-Higgs production. 
The one on the left, where two Higgs bosons are produced via an off-shell Higgs ($H^*$), is the only process directly influenced by the value of the trilinear Higgs self-coupling. The different kinematics of the $HH$ events coming from $H^*$ is exploited to extract the value of $\lambda_3$.
The strategy employed to estimate the uncertainty on the trilinear Higgs self-coupling is as follows:
\begin{itemize}
    \item Samples of double-Higgs events are generated with WHIZARD for different values of $\kappa_{\lambda_3}=\lambda_3/\lambda_{3}^{SM}$: 0.2, 0.4, 0.6, 0.8, 0.9, 1.0, 1.1, 1.2, 1.4, 1.6, 1.8. Subsequently, the samples are simulated and reconstructed with the procedure described in Sec.~\ref{sec:HH}. The $\kappa_{\lambda_3}=1$ case represents the SM signal.
    Tab.~\ref{tab:kappa_values} shows the expected number of $HH$ events for different $\kappa_{\lambda_3}$ hypotheses;
    \item Two MLPs are trained independently:
    \begin{enumerate}
        \item The first MLP is the same as the one used for the $HH$ cross section determination in Sec.~\ref{sec:HH} to separate the SM signal ($\kappa_{\lambda_3}=1$) from the physics background.
        \item The other MLP is trained on four variables to differentiate between the $\mu^+ \mu^- \to HH \nu_\mu \bar{\nu}_\mu \to b\bar{b}b\bar{b} \nu_\mu \bar{\nu}_\mu$ process,
        where the Higgs pair is produced exclusively via an off-shell $H^*$ 
        and the other two contributions.
        The rationale behind this approach is to construct a MLP that leverages the kinematic properties specific to the production of Higgs pairs via an off-shell Higgs boson, which is indicative of the trilinear Higgs self-coupling. This allows for the separation of this process from other double-Higgs production processes.
        The variables that allow the best separation 
        are: the angle between the two Higgs boson momenta in the laboratory frame, the angle between the highest-$p_{T}$ jet momenta of each pair with respect to the $z$ axis, and the helicity angle of the two Higgs boson candidates.
    \end{enumerate}
    
 \item The scores of the two MLPs for the considered samples have been arranged in 2-dimensional histograms. To obtain the expected data distribution, 2-dimensional templates of the signal and background components are built for each $\kappa_{\lambda_3}$ hypothesis. Signal and background yields calculated in Tabs.~\ref{tab:HH_negl_mistag} and \ref{tab:kappa_values} are used to weight the templates.
 \item Pseudo-datasets are generated with the total 2D template for the $\kappa_{\lambda_3}=1$ hypothesis. For each pseudo-experiment, the likelihood difference $-\Delta \mathrm{log} (L)$ is calculated as a function of $\kappa_{\lambda_3}$  by comparing the pseudo-data distribution to the $\kappa_{\lambda_3}$ templates.
 \item The log-likelihood profile has been fitted with a polynomial function of fourth degree. The uncertainty on $\kappa_{\lambda_3}$ at 68\% C.L. is estimated as the interval around $\kappa_{\lambda_3}=1$ where the fitted polynomial has a value below 0.5.
\end{itemize}



As shown in Tab.~\ref{tab:kappa_values}, the expected production cross sections for $\kappa_{\lambda_3}>1$ are very similar.  
Therefore, distinguishing among them would require optimizing the jet reconstruction algorithm with lower thresholds, which would result in very high computational times.
The result obtained with the current algorithms, does not provide an adequate representation of the true potential of MuC for this measurement. 
Nevertheless, it is possible to assess the impact of better performing algorithms on the measurement, as discussed in the following:

\begin{itemize}
\item With dedicated machine-learning algorithms that exploit the jet sub-structure, it is possible to identify $b$-jets with a performance comparable to that of hadron colliders. Such  algorithms are currently employed at LHC to enhance the performance of jet flavor identification both in the central region of $pp$ collisions, like the DeepJet algorithm~\cite{deepjet} at CMS, and in the forward region at LHCb~\cite{dnn}. 
At a MuC, $b$-jets from $HH$ are expected to be produced mainly in the forward region, which resembles the phase space covered by LHCb.
Therefore, the LHCb tagging capabilities are assumed for the MuC expectation: the $b$-tagging efficiency and $c$-mistag are taken from Ref.~\cite{dnn}, while light jet mistag is assumed negligible, since its value is really low (in the order of 0.1$\%$ in LHCb data). 
A working point for the $HH$ analysis has been determined by maximizing the significance ($S/\sqrt{S+B}$) of $HH$ events across various configurations of LHCb $b$-tagging and $c$-mistag performance. 
The optimal configuration, yielding the highest significance, requires: 3 tagged jet, a total $b$-tagging efficiency of 76$\%$ and a $c$-mistag rate of 20$\%$. Simulated events for $HH$ signal and background processes are re-weighted according to these efficiencies.
\item  Several developments are currently underway to minimize the impact of the BIB on multi-jets final states reconstruction performance aiming to reduce it to a negligible level.
For example, the number of BIB hits in the tracker system could be reduced by applying a cutting-edge clustering algorithm capable of identifying and rejecting hits associated with low-momentum BIB particles, which tend to deposit more ionization energy. This strategy aims to decrease the occurrence of spurious tracks, thereby enhancing the overall tracking performance. 
The effect of the calorimeter energy threshold is studied by reducing it from 2 MeV to 200 KeV. The improvement in the jet energy resolution, demonstrated in Ref.~\cite{epjc}, will impact the $HH$ analysis. The invariant mass distribution of the leading Higgs candidate in $HH$ events decaying to $q_h \bar{q}_h q_h \bar{q}_h$ events under the two thresholds hypothesis, both without the BIB overlay, is shown in Fig.~\ref{inv_mass_wo_BIB} (left).
The improved separation between the Higgs and $Z$ boson resonances with respect to what has been obtained in Fig.~\ref{fig:hbb_fit} is evident. The jet momentum resolution in the case without the BIB overlay is in the order of 10\%, while the resolution with the BIB is of about 15\%.
This 10\% resolution is used in the analysis to evaluate the trilinear coupling precision.
\end{itemize}

The likelihood scan obtained with the previous assumptions on the tagging efficiencies and jet momentum resolution is shown in Fig.~\ref{inv_mass_wo_BIB} (right). The uncertainty on the trilinear Higgs self-coupling is found to be $0.81<\kappa_{\lambda_3}<1.44$ at 68 $\%$ confidence level. It could be noticed that the upper bound is higher since the number of signal events as a function of $\kappa_{\lambda_3}$ varies slowly for $\kappa_{\lambda_3}>1.0$, as can be seen in Tab.~\ref{tab:kappa_values}.

The result obtained here can be compared with the values found in~\cite{3tevphysics,epjc}; for first minimum with the same integrated luminosity of $\mathcal{L}=1$ ab$^{-1}$, it is reported $0.73 < \kappa_{\lambda_3} < 1.35$. 
 
\section{Systematic uncertainties}
\label{sec:sys}

The errors quoted in the previous sections include only the statistical uncertainties, reflecting the level of precision targeted with one experiment at 1 ab$^{-1}$ of data collected at a center-of-mass energy of 3 TeV. This study is crucial for determining the detector requirements in the new background environment.

A precise evaluation of the systematic uncertainties would require knowledge of:
\begin{itemize}
\item [-] the BIB for a given interaction region's optical configuration, including the nozzle shape and material, which can influence the number of produced particles and their time, energy, and spatial distributions;
\item[-] the detector implementation, such as the positioning of the detector services and their impact on the material distribution;
\item[-] the combined effect of these factors that determines the number of particles reaching the detector and their kinematic properties.
\end{itemize}
This is particularly challenging for high-energy muon collisions, where there is little prior experience. Nevertheless, a discussion based on simulations and expertise from other experiments is provided to present the potential sources of systematic uncertainties commonly considered.


\begin{description}

\item[\textbf{Luminosity:}]
The determination of the absolute integrated luminosity for muon collisions at $\sqrt{s}=3$ TeV has not yet been studied. However, an evaluation of potential precision has been performed for $\sqrt{s}=1.5$ TeV ~\cite{annual}. This study is based on the measurement of the number of $\mu^+\mu^-\rightarrow \mu^+\mu^-$ (muon Bhabha) events at large angles, $30^\circ <\theta < 150^\circ$, to avoid the impact of the nozzles. The statistical precision on the number of events is at the permille level. To achieve similar precision for $\mathcal{L}$, the theoretical cross section must be known with equivalent accuracy, which is not yet available for muon interactions at this energy. If the Muon Collider is constructed, theoretical predictions are expected to reach the required precision. Therefore, assuming an accuracy at the percent level is reasonable.

\item[\textbf{Jet energy scale:}]
After the calorimeter calibration, the remaining energy correction for jets originating from Higgs boson decays will be evaluated once samples of heavy-flavor and light-flavor jets become available. 
For heavy-flavor jets, the procedure will rely on the reconstruction of the $Z \rightarrow b\bar b$ invariant mass, as $b$-jets are produced with momenta similar to those from Higgs boson decays. The expected number of $Z \rightarrow b\bar b$ events with the same requirements of the analysis in Sec. \ref{sec:h2bb}, approximately 40k, should allow for a precision better than $0.5\%$ for the cases presented. Using this decay channel for calibration will also improve the energy resolution of $b$-jets.

\item[\textbf{Flavor tagging:}]
In the presented studies, jets originating from $b$-, $c$-, and light-quarks are identified based on the properties of the secondary vertex reconstructed within the jet itself. More advanced algorithms \cite{deepjet,dnn}, can be implemented, which is why, for multi-jet events, the performance observed in the LHCb experiment is assumed to represent a more realistic scenario. The systematic uncertainty on jet tagging on absolute efficiencies and on the kinematical variables will be evaluated using the tag-and-probe method applied to $Z \rightarrow b\bar b$ events in addition to the other methods in use at LHC. Assuming a $0.5\%$ uncertainty, which is a very conservative estimate, the error on the $H \rightarrow b\bar b$ cross section is $0.5\%$.

\end{description}

\section{Lesson learned: requirements for Higgs measurements}
\label{sec:requirements}

\begin{table*}[t]
    \centering
    \caption{Detector and reconstruction algorithm performance requirements for Higgs physics at the Muon Collider, determined with the detailed detector simulation studies. The values have been obtained for objects with $p_T$ of about 100 GeV, which is the average transverse momentum of the particles from the 2-body decay of the Higgs boson with $\sqrt{s}=3$ TeV. The first set of requirements is intended for single Higgs boson measurements, while more stringent requirements are presented for the trilinear coupling determination with the four-$b$-jet final state.}
    \begin{tabular}{c|c}
    \hline
         \textbf{Object} & Requirements \\
    \hline
    \hline
       muons &  $\frac{\Delta p_T}{p_T} = 0.4 \%$\\
      \hline
       photons & $\frac{\Delta E}{E} = 3\%$ \\
      \hline
       jets & $\frac{\Delta p_T}{p_T} = 15 \%$ \\
      \hline
       $b$-jets & $\frac{\Delta p_T}{p_T} = 15 \%$\\
                        &  $b$ efficiency = 60 \%\\
                        &  $c$ mistag = 20 \%\\
      \hline
      \hline
       $b$-jets & $\frac{\Delta p_T}{p_T} = 10 \%$ \\
            (in multi-jet event) &  $b$ efficiency = 76 \%\\
        &  $c$ mistag = 20\%\\
      \hline
      \hline
    \end{tabular}
    \label{tab:requirements}
\end{table*}

\begin{figure}[b]  
    \centering
    \includegraphics[width=0.48\textwidth]{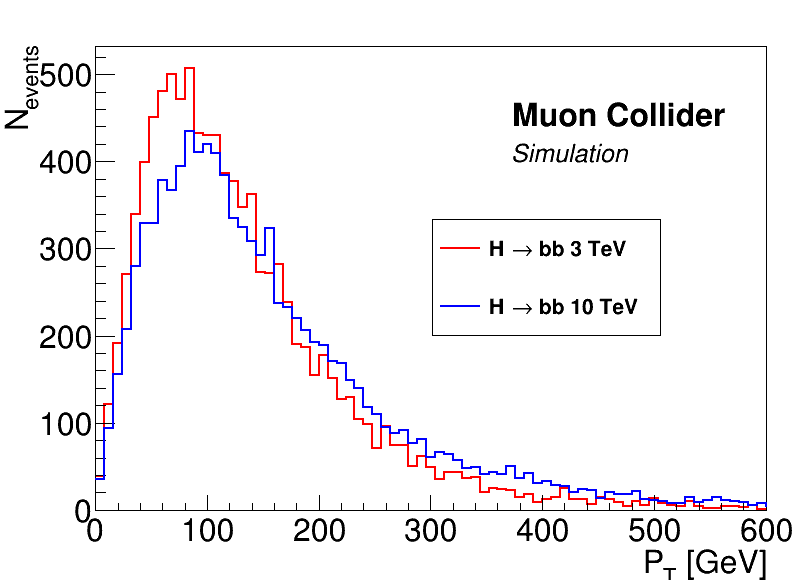}
    \caption{Transverse momenta of $b$-quarks from the $H \rightarrow b \bar{b}$ decay, at $\sqrt{s}=3$ and $\sqrt{s}=10$ TeV muon-antimuon collisions, determined with Madgraph~\cite{det_eps}.
    \label{fig:10tev}}
\end{figure}

The Higgs studies presented in this paper can serve as a guideline to establish the requirements for designing a detector for $\sqrt{s}=10$ TeV and to improve the reconstruction algorithms of the physics objects at any center-of-mass energy.
In fact, the transverse momentum distributions of $b$-quarks from $H \to b \bar{b}$ decays at $\sqrt{s}=3$ and $\sqrt{s}=10$ TeV are very similar, with a peak around $p_{\mathrm{T}} \sim 100$ GeV as shown in Fig.~\ref{fig:10tev}.

Sec.~\ref{sec:parametric} demonstrates that the results obtained for a $\sqrt{s}=3$ TeV MuC with the detailed detector simulation are close to those determined with target performance.
%
Sec.~\ref{sec:higgsxs} has shown how the detector and reconstruction algorithms, discussed in Sec.~\ref{sec:reconstruction}, are sufficient to control the BIB impact on the performance 
and allow to achieve the precision on single Higgs cross section, $H \to b \bar{b}$, below $1\%$.
On the other hand, Sec.~\ref{sec:trilinear} has shown how several improvements on jet reconstruction and identification are needed to improve the analysis performance on multi-jets final state, therefore the precision on the Higgs trilinear coupling.

In light of the previous considerations, the requirements listed in Tab.~\ref{tab:requirements} are determined as follows:
\begin{itemize}
\item For muons, photons, jets, and $b$-jets the requirements on the resolutions are taken from the reconstruction performance described in Sec.~\ref{sec:reconstruction}, and evaluated for objects with $p_{\mathrm{T}}=100$ GeV. The value of $p_{\mathrm{T}}=100$ GeV is chosen based on the peak position in the $p_{\mathrm{T}}$ distribution of $b$ quarks from Higgs boson decays (Fig.~\ref{fig:10tev}). For $b$-jets the identification efficiencies and the $c$-mistag rate are also quoted. 
These requirements are sufficient to achieve a statistical sensitivity on the Higgs boson production cross sections comparable to those obtained with the assumed target performance.
\item The efficiencies and resolutions described in Sec.~\ref{sec:reconstruction} are not sufficient to achieve the precision on $\lambda_3$ obtainable with the assumed target performance. This will be true for any multi-jet final state. Therefore, for this event topology, the requirements, in particular for $b$-jet identification and $p_{\mathrm{T}}$ resolution, are taken from the performance described in Sec.~\ref{sec:trilinear} that is based on reasonable assumptions. 
%
\end{itemize}
The performance requirements are summarized in Tab.~\ref{tab:requirements}.


\section{Summary and Outlook}
\label{sec:conclusions}

The results presented in this paper demonstrate that, at a $\sqrt{s}=3$ TeV MuC, state-of-the-art detector technologies and reconstruction algorithms enable high-precision measurements of Higgs cross sections, including that for double Higgs production and the Higgs trilinear self-coupling. 
The sensitivities have been estimated using a detailed detector simulation that includes the BIB~\cite{mdi_eps}. For each analysis, the Higgs boson signal events and their corresponding physics backgrounds were fully simulated and reconstructed (Sec.~\ref{sec:reconstruction}).  
The statistical sensitivities on the Higgs cross section measurements (Sec.~\ref{sec:higgsxs}) for the considered Higgs decay channels ($H\rightarrow b\bar b$, $H\rightarrow W W^*$, $H\rightarrow Z Z^*$, $H\rightarrow \mu^+ \mu^-$, and $H\rightarrow \gamma\gamma$) are consistent with those obtained using target detector simulations without the BIB. This consistency underscores the capability of the detector configuration and reconstruction algorithms to effectively manage the effects of the BIB.

This indicates that the 3 TeV center-of-mass energy is a strong starting energy for studying the Higgs properties  with muon collisions, though it presents limitations for measuring the Higgs width, as indicated by Refs.~\cite{Forslund:2023reu, Li:2024joa} and by the analysis presented here in Sec.~\ref{sec:width}. 

The precision on the Higgs trilinear self-coupling, discussed in Sec.~\ref{sec:double}, was evaluated under the reasonable assumption that a $b$-jet identification performance comparable to that of LHCb can be achieved. This result is compatible with that obtained with parametric studies in Refs.~\cite{3tevphysics,epjc}. 



Another significant outcome of this paper is the determination of detector requirements for Higgs physics (Sec.~\ref{sec:requirements}), derived from detailed analyses that utilize various physics objects. These requirements, listed in Tab.~\ref{tab:requirements}, have been found to hold also at higher center-of-mass energies, as for example $\sqrt{s}=10$ TeV, and can serve as a basis for designing a detector suitable to these energy levels.
Indeed, the kinematic distributions of the Higgs boson are quite similar at both 3 TeV and 10 TeV center-of-mass energies~\cite{det_eps}.
In this scenario, with the BIB at $\sqrt{s}=10$ TeV expected to be similar to that at $\sqrt{s}=3$ TeV~\cite{mdi_eps}, the potential of muon collisions at this higher energy could be fully realized. For example, MuC running at $\sqrt{s}=10$ TeV for 5 years with an integrated luminosity of 10 ab$^{-1}$, could achieve an uncertainty of approximately 4\% on the Higgs trilinear self-coupling, representing one of the best precisions attainable by any future collider.

\section*{Acknowledgments}
We are grateful to the International Muon Collider Collaboration and the US Muon Accelerator Program for their support. In particular, we thank Anthony Badea, Tova Holmes, Federico Meloni, and Andrea Wulzer for their valuable comments that helped improve the paper.
We acknowledge the financial support of the Italian National Institute for Nuclear Physics (INFN), the University of Padua, and the European Organization for Nuclear Research (CERN).
This work was supported by the European Union’s Horizon 2020 and Horizon Europe Research and Innovation programs through the Marie Sk\l{}odowska-Curie RISE Grant Agreement No. 101006726, the Research Infrastructures INFRADEV Grant Agreement No. 101094300, and the EXCELLENT SCIENCE - Research Infrastructures Research Innovation Grant Agreement No. 101004761.

\bibliography{sn-bibliography}

\newpage

\appendix

\section{Appendix: list of Monte Carlo samples}
\label{app}

Table~\ref{tab:MC} summarizes all Monte Carlo samples used in this paper. It lists the simulated processes, the generators employed, and the generator-level cuts.
Details on the event generation are provided in Sec.~\ref{sec:generators}.
The generation requirements are applied to $P_{\mathrm{T}}^i$, the transverse momentum with respect to the $z$ axis (where $i$ is the particle type), to the invariant mass of two particles $M(i,j)$, to the pseudorapidity $\eta^i$ or to the distance between two particles in the $\eta$-$\phi$ space, $\Delta R(i,j) = \sqrt{(\Delta \eta)^2+(\Delta \phi)^2}$, where $\phi$ is the angle with respect to the $x$ axis in the $x$-$y$ plane.

\begin{table*}[t!]
    \centering
    \caption{Summary of the generated samples indicating the Monte Carlo generator used and the kinematical requirements applied to the final-state particles. $M$ represents the invariant mass of the objects shown in parentheses. The symbol $q$ stands for $u$, $d$, $s$, $c$, or $b$ quarks, while $\ell$ stands for electron, muon or tau. Symbol $\nu$ stands for electron, muon or tau neutrino. 
    If $q\bar{q}$ is present, all possible combination of quark flavours, even different flavours, are considered by WHIZARD/Madgraph.
    If the Higgs boson is not indicated in the process, Yukawa couplings are switched off.}
    \begin{tabular}{l| c c }
    \hline
    \hline
         \textbf{Process}                   &  Generator  & Kinematical requirements \\
    \hline
       $ \mu^+\mu^- \to H\nu_\mu\bar{\nu}_\mu;\: H\to b \bar{b} $ & WHIZARD   & - \\
       $ \mu^+\mu^- \to H\nu_\mu\bar{\nu}_\mu;\: H\to c \bar{c} $ & WHIZARD   & - \\
       $ \mu^+\mu^- \to H\nu_\mu\bar{\nu}_\mu;\: H\to g g $ & WHIZARD   & - \\
       $ \mu^+\mu^- \to H\nu_\mu\bar{\nu}_\mu;\: H\to W W^*\to q \bar q \mu\nu_\mu $ & WHIZARD  &  $M(q,\bar q)> 10$ GeV  \\
       $ \mu^+\mu^- \to H\nu_\mu\bar{\nu}_\mu;\: H\to W W^*\to q \bar q q \bar q $ & WHIZARD  &  $M(q,\bar q)> 10$ GeV  \\
       $ \mu^+\mu^- \to H\nu_\mu\bar{\nu}_\mu;\: H\to \tau^+ \tau^-$ & WHIZARD & - \\
       $ \mu^+\mu^- \to H\nu_\mu\bar{\nu}_\mu;\: H\to\mu^+\mu^-$        & MadGraph   &  -  \\
       $ \mu^+\mu^- \to H H\nu_\mu\bar{\nu}_\mu;\: H\to b \bar{b}, H\to b \bar{b} $ & WHIZARD & -\\
       $ \mu^+\mu^- \to H\mu^+\mu^-;\: H\to W W^*\to q \bar q \mu\nu_\mu$ & WHIZARD   &  $M(q,\bar q)> 10$ GeV \\
       $ \mu^+\mu^- \to H\mu^+\mu^-;\: H\to\mu^+\mu^-$ & MadGraph   &  $M(\mu,\mu)> 100$ GeV \\
       $ \mu^+\mu^- \to H\nu_\mu\bar{\nu}_\mu;\: H\to\gamma\gamma$ & MadGraph     & $P_{\mathrm{T}}^{\gamma}>10$ GeV, $|\eta^\gamma| < 2.436$\\
       $ \mu^+\mu^- \to Z H\nu_\mu\bar{\nu}_\mu;\: Z\to\mu^+ \mu^-, H\to b \bar{b}$ & MadGraph & $P_{\mathrm{T}}^{\mu}$ > 1 GeV, $|\eta^{\mu}|$ < 3\\
       										&	     & $P_{\mathrm{T}}^q$ > 5 GeV, $|\eta^q|$ < 3 \\
										&	     & $\Delta R(q,\bar{q})$ > 0.2, $\Delta R(q,\mu) > 0.05$\\
    $ \mu^+\mu^- \to H Z;\: H\to W W^*\to q \bar q \mu\nu_\mu, Z\to \mu^+\mu^-$ & WHIZARD   & $M(q, \bar{q})> 10$ GeV  \\
    $ \mu^+\mu^- \to H q \bar{q} \nu\bar{\nu};\: H\to b \bar{b} $ & WHIZARD   & $M(q, \bar{q})> 10$ GeV \\
     $ \mu^+\mu^- \to Z Z\nu\bar{\nu};\: Z\to\mu^+ \mu^-, Z\to q \bar{q}$ & MadGraph & $P_{\mathrm{T}}^{\mu}$ > 1 GeV, $|\eta^{\mu}|$ < 3\\
										&	     & $P_{\mathrm{T}}^q$ > 5 GeV, $|\eta^q|$ < 3 \\
										&	     & $\Delta R(q,\bar{q})$ > 0.2, $\Delta R(q,\mu) > 0.05$\\
       $ \mu^+\mu^- \to W^{\pm} Z \mu \nu;\: W^{\pm}\to q \bar{q}, Z\to\mu^+ \mu^-$ & MadGraph & $P_{\mathrm{T}}^{\mu}$ > 10 GeV \\
	    									&	     & $P_{\mathrm{T}}^q$ > 5 GeV, $|\eta^q|$ < 3 \\
										&	     & $\Delta R(q,\bar{q})$ > 0.2, $\Delta R(q,\mu) > 0.05$\\
       $ \mu^+\mu^- \to q \bar q \nu\bar{\nu}$       &  WHIZARD   &   $M(q, \bar{q})> 10$ GeV\\
       $ \mu^+\mu^- \to q \bar q \ell\ell$       &  WHIZARD   &   $M(q, \bar{q})> 10$ GeV\\
       $ \mu^+\mu^- \to q \bar q \ell \nu $       &  WHIZARD   &    $M(q, \bar{q})> 10$ GeV \\
       $ \mu^+\mu^- \to q \bar{q} q \bar{q} \nu \bar{\nu}$       &  WHIZARD   &    $M(q, \bar q)> 10$ GeV \\

       $ \mu^+\mu^- \to q \bar{q} q \bar{q} \ell \ell$       &  WHIZARD   &    $M(q, \bar q)> 10$ GeV \\
       $ \mu^+\mu^- \to q \bar{q} q \bar{q} \ell \nu$       &  WHIZARD   &    $M(q, \bar q)> 10$ GeV \\
       $ \mu^+\mu^- \to q \bar{q} q \bar{q}$       &  WHIZARD   &    $M(q, \bar q)> 10$ GeV \\
       $ \mu^+\mu^- \to \nu_\mu\bar{\nu}_\mu \mu^+\mu^- q \bar{q}$ & WHIZARD & 10 GeV < $M(q,\bar{q})$ < 150 GeV \\
									&	 & $|\eta^{\mu}|$ < 2.5, $|\eta^q|$ < 2.5 \\
									&	 & $P_{\mathrm{T}}^{\mu}$ > 5 GeV, $P_{\mathrm{T}}^q$ > 5 GeV \\
									&	 & $\Delta R(q,\bar{q})$ > 0.3 \\
       $ \mu^+\mu^- \to \mu^+\mu^-\nu_\mu\bar{\nu}_\mu$    & MadGraph   &  $ 100 < M(\mu,\mu)< 150$ GeV, \\
                                                   &            & $|\eta^\mu|<2.66$ \\
       $ \mu^+\mu^- \to \mu^+\mu^-\mu^+\mu^-$ & MadGraph   &  $ 100 < M(\mu,\mu)< 150$ GeV, \\
                                                   &            & $|\eta^\mu|<2.66$ \\
       $ \mu^+\mu^- \to t\bar{t};\: t\to W b, W\to\mu\nu_\mu$  & MadGraph   &  $ M(\mu,\mu)> 50$ GeV \\
       $ \mu^+\mu^- \to \gamma\gamma\nu\bar{\nu}$& Madgraph   & $P_{\mathrm{T}}^{\gamma}>10$ GeV, $|\eta^\gamma| < 2.436$\\
       $ \mu^+\mu^- \to \ell\ell \gamma$& Madgraph   & $P_{\mathrm{T}}^{\gamma}>10$ GeV, $|\eta^\gamma| < 2.436$\\
       $ \mu^+\mu^- \to \ell\ell \gamma\gamma$& Madgraph   & $P_{\mathrm{T}}^{\gamma}>10$ GeV, $|\eta^\gamma| < 2.436$\\
       $ \mu^+\mu^- \to \gamma\gamma$& Madgraph   & $P_{\mathrm{T}}^{\gamma}>10$ GeV, $|\eta^\gamma| < 2.436$\\
    \hline
    \hline
    \end{tabular}
    \label{tab:MC}
\end{table*}

\end{document}